\documentclass{autart}

\usepackage{dsfont}
\usepackage{bbm}
\usepackage{mathrsfs}
\usepackage{natbib}
\usepackage{color}
\usepackage{graphics} 
\usepackage{latexsym, amsmath,theorem, amsfonts, amssymb,graphicx,array,tabularx,algorithm}
\usepackage{booktabs}
\usepackage{subfigure}
\usepackage{caption}
\usepackage{algpseudocode}
\usepackage{tikz}
\usepackage{pgfplots}
\pgfplotsset{every axis/.append style={font=\scriptsize}}
\usetikzlibrary{calc,plotmarks,shapes}

\theoremstyle{theorem}
\newtheorem{theorem}{Theorem}
\newtheorem{lemma}{Lemma}

\newtheorem{proposition}{Proposition}
\newtheorem{remark}{Remark}

\newtheorem{example}{Example}
\newtheorem{definition}{Definition}

\newtheorem{problem}{Problem}

\newcommand{\R}{{\rm  I\kern-2pt R}}
\renewcommand{\Re}{{\rm  I\kern-2pt R}}

\newcommand{\ie}{{\it i.e.}}

\newcommand{\argmax}{\arg\max}

\newcommand{\card}[1]{|#1|}
\newcommand{\tr}[1]{\mathrm{Tr}\left[#1\right]}
\newcommand{\bb}[1]{\boldsymbol{#1}}

\newcommand{\I}{{\bb{\rm I}}}
\newcommand{\1}{{\textbf{1}}}

\newcommand{\hlb}[1]{\textcolor[rgb]{0.00,0.00,0.00}{#1}}

\begin{document}
\begin{frontmatter}

\title{Optimal Sensor Scheduling for Multiple Linear Dynamical Systems}


\author[NTU,HKUST]{Duo Han}\ead{dhanaa@ntu.edu.sg},
\author[HKUST,KTH]{Junfeng Wu}\ead{junfengw@kth.se},
\author[SDU]{Huanshui Zhang}\ead{hszhang@sdu.edu.cn.},
\author[HKUST]{Ling Shi}\ead{eesling@ust.hk}

\address[NTU]{School of Electrical and Electronic Engineering, Nanyang Technological University, Singapore.}
\address[HKUST]{Department of Electronic and Computer Engineering, Hong Kong University of Science and Technology, Hong Kong.}
\address[KTH]{ACCESS Linnaeus Center, School of Electrical Engineering, Royal Institute of Technology, Stockholm, Sweden.}
\address[SDU]{School of Control Science and Engineering, Shandong University, Jinan, 250061, Shandong, P. R. China.}


\begin{abstract}
We consider the design of an optimal collision-free sensor schedule for a number of sensors which monitor different linear dynamical systems correspondingly. At each time, only one of all the sensors can send its local estimate to the remote estimator. A preliminary work for the two-sensor scheduling case has been studied in the literature. The generalization into multiple-sensor scheduling case is shown to be nontrivial. We first find a necessary condition of the optimal solution provided that the spectral radii of any two system matrices are not equal, which can significantly reduce the feasible optimal solution space without loss of performance. By modelling a finite-state Markov decision process (MDP) problem, we can numerically search an asymptotic periodic schedule which is proven to be optimal. From a practical viewpoint, the computational complexity is formidable for some special system models, e.g., the spectral radii of some certain system matrices are far from others'. Some simple but effective suboptimal schedules for any systems are proposed. We also find a lower bound of the optimal cost, which enables us to quantify the performance gap between any suboptimal schedule and the optimal one.
\end{abstract}

\end{frontmatter}

\section{Introduction}\label{section:introduction}
Wireless sensor networks (WSNs) have been playing an irreplaceable role in modern control and estimation systems such as target tracking and localization \citep{martinez2006optimal,bishop2010optimality,hu2010nonlinear,hoang2014sensor,bai2015robust}. Typically in most real applications, sensors are sparsely placed to monitor one or multiple dynamical processes of interest and noisy measurements collected will be sent back to the estimator for generating the state estimate further used by the controller. In most cases, the sensors often share a common wireless channel to complete specific estimation tasks \citep{savage2009optimal,mo2011sensor,shi2012scheduling,vitus2012efficient,huber2012optimal,han2013event,mo2014infinite,zhao2014optimal}. As the number of sensors increases, the lack of an efficient network scheduling protocol will lead to severe information loss and thus poor estimation quality. Therefore, the problem of scheduling multiple sensors over a shared channel to achieve good estimation performance arises. For example, contention-based medium access control (MAC) has been studied in \cite{chen2010random}. The authors proposed a novel medium access method achieving the Nash equilibrium of a random access game.

Rather than resolve the potential contention during transmission, in this article we focus on the collision-free scheduling policies where only one single sensor among a network of sensors can transmit its data packet at each time. The problem of collision-free scheduling problems are mainly classified into two categories in terms of how many dynamical systems are being monitored: 1) scheduling the sensors observing one single common system \citep{alriksson2005sub,arai2007optimal,vitus2012efficient,shi2013approximate,zhao2014optimal}; 2) scheduling the sensors observing different systems \citep{savage2009optimal,shi2012scheduling,lin2013scheduling,Li2015173}. As an example of the single-system type, the problem of choosing which
sensor should operate at each time-step to minimize a weighted function of the error covariances of the
state estimates was considered in \cite{vitus2012efficient}. \hlb{Based on a condition on when an initial schedule is not part of the optimal schedule, both algorithms for searching optimal and suboptimal solutions for finite-horizon problems were proposed.} The result is extended in \cite{zhao2014optimal} for an infinite-horizon sensor scheduling problem. It has been proved that under some mild conditions, both the optimal infinite-horizon average-per-stage cost and the corresponding optimal sensor schedules are not affected by the covariance matrix of the initial state. Another interesting finding in \cite{zhao2014optimal} is that the optimal estimation cost can be approximated arbitrarily closely by a periodic schedule. The aforementioned works focus on the scheduling problem where multiple sensors measure a single system of common interest. More related works of the single-system type can be found therein \citep{shi2013approximate,orihuela2014periodicity,jawaid2015submodularity} .

\begin{figure}
  \centering
  \includegraphics[height=2.4in]{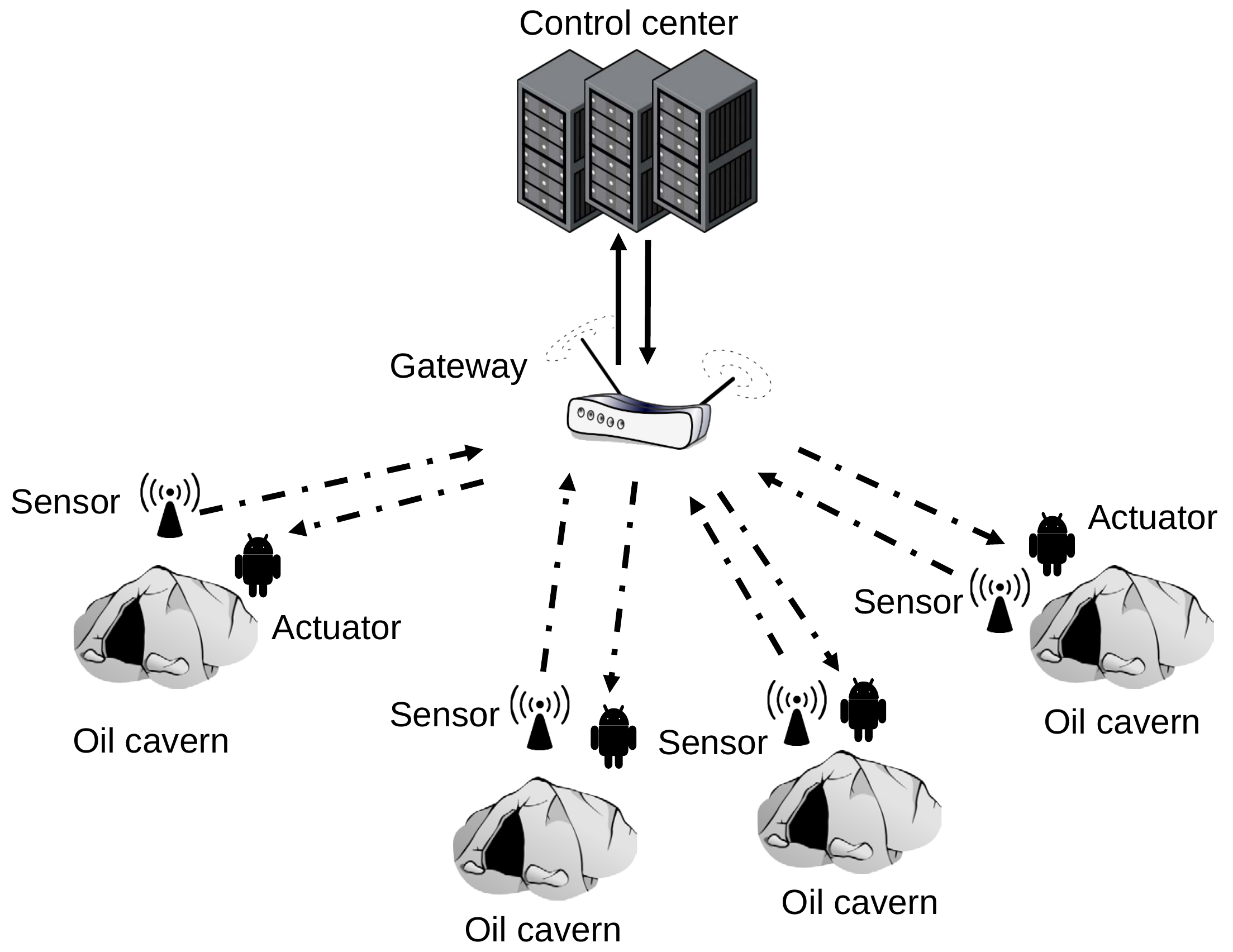}\\
  \caption{Underground petroleum storage.}\label{fig:motivation}
\end{figure}

Compared with the single system case, not enough research efforts have been put in the case of different sensors monitoring different systems, which is widely encountered in practice. A simple motivating example is underground petroleum storage using \emph{Wireless}HART technology in \cite{song2008wirelesshart} (see Fig. \ref{fig:motivation}). Underground salt caverns are often used for crude oil storage. Brine and crude oil flowing in both directions are measured by the sensors and reported to the estimation and control center through a gateway device. The control center targets for maintaining a certain pressure inside the caverns. Apparently, each sensor competes with the others for the gateway access to achieve their own goal. Therefore, a schedule for optimizing a cost function consisting of all sensors' benefits is desirable. Different performance metrics may result in different optimal schedules. The authors studied the multiple sensors scheduling for different scalar Gauss-Markov systems over a finite horizon in \cite{savage2009optimal}. They considered the optimality in terms of terminal estimation error covariance. The optimal policy is to schedule the transmissions in the end of the horizon in a specified order. However, the terminal error covariance is only suitable for modelling finite horizon problem. A metric of the trace of averaged estimation error covariance is adopted in \cite{shi2012scheduling}. The authors studied two multi-dimensional systems over an infinite horizon and proposed an explicit optimal sensor schedule for this two-sensor case. The approach, however, cannot be generalized into the case of three sensors or more, which limits its scope. Another similar problem where one sensor is scheduled to monitor multiple systems was studied in \cite{lin2013scheduling}. The problem is same with multiple sensor scheduling problem in nature. The authors proposed algorithms to search a schedule such that the error covariance of each system is bounded by some constant matrix. A stochastic solution was also discussed in \cite{Li2015173}.

In this work we consider the problem of designing a deterministic collision-free optimal schedule for the sensors monitoring different linear dynamical systems over an infinite horizon. Each sensor collects the noisy measurement of an individual system and generates a local estimate at each time instant. Only one of these sensors is allowed
to access the shared channel, sending its data to the remote estimator, for transmission media access control. Our goal is to design an optimal sensor schedule to minimize the trace of the averaged estimation error covariance of all processes.
We restrict our attention to a type of
 {time-based} sensor transmission schedule that specifies the scheduling decisions for all sensors according to time and aim to generalize the results in two-sensor case \cite{shi2012scheduling} in the framework of multiple sensors scheduling. This extension, as we will show, is not trivial. The simpleness of the optimal schedule proposed in \cite{shi2012scheduling}, e.g., in each period one transmission of one sensor followed by several transmissions of the other one, is mainly because of mutual exclusiveness of the two sensors. When the number of the sensors is not less than three, the periodicity and the structure of the optimal schedule remain unclear.

 To tackle the optimal multiple sensor scheduling problem, we first curtail the feasible solution space by proving a necessary condition of the optimal schedule: the maximum off-duty duration of each sensor is upper bounded by a finite number provided that the spectral radii of any two system matrices are not equal. We then address the optimal multiple sensor scheduling problem by modelling a finite-state Markov decision process (MDP). However, the complexity of solving the MDP increases dramatically if the derived upper bound depending on the system model is large. Therefore, we also propose some suboptimal schedules for practical use and we can effectively quantify the performance gap between the optimal schedule and the suboptimal ones by giving an upper bound on that gap. The result in the two-sensor case \cite{shi2012scheduling} is shown to be a special case of our study.
The contributions of this work are summarized as follows:
\begin{enumerate}
  \item \hlb{We study the optimal collision-free sensor scheduling protocol for multiple sensors monitoring different linear dynamical systems. We find two necessary conditions for the optimal schedule, e.g., the boundedness of off-duty durations and the uniformity of transmissions, which can be used to help identify the schedule which is not optimal.}
  \item By exploiting the necessary conditions, we can solve the optimal scheduling problem by modelling it as a finite-state MDP problem. The existence of an optimal solution is proved. We also show that the optimal schedule converges to be periodic.
  \item To circumvent the high computational complexity of solving the MDP problem when the number of the states is huge, we propose several suboptimal schedules for practical use. We also find a lower bound for the optimal cost and thus we can obtain an upper bound on the performance gap between the suboptimal schedules and the optimal schedule.
\end{enumerate}

The remainder of the paper is organized as follows. Section~\ref{section:problem-setup} describes the problem of interest. How to search and identify an optimal schedule is presented in Section \ref{section:optimal}. More results on suboptimal schedules are discussed in Section \ref{section:suboptimal}. Concluding remarks are given in Section \ref{section:conclusion}.

\textit{Notations}: All vectors and matrices are named in boldface while scalars are not. \hlb{$\mathbb Z_+$} is the set of nonnegative integers and \hlb{$\mathbb Z_{++}:=\mathbb Z_+\backslash \{0\}$}.  For $x\in\mathbb R$, $\lfloor x\rfloor$ is the largest integer that is not larger than $x$ and
$\lceil x\rceil$ is the smallest one that is not less than $x$. The nonnegative integer $k$ is the time index. $\mathbb{S}_{+}^{n}$ is the set of $n$ by $n$ positive semi-definite  matrices. When $\bb X \in \mathbb{S}_{+}^{n}$, we write $\bb X \geq \bb 0$; when $\bb X$ is positive definite, we write $\bb X > 0$. $\bb X^{\top}$ denotes the transpose of the matrix $\bb X$. $\mathrm{Tr}[\cdot]$ denotes the trace of a matrix. $\mathbb{E}[\cdot]$ denotes the expectation of a random variable. For a vector $\bb x\in \mathbb{R}^n$, we use
$\bb x[i]$ to denote the $i$th entry of $\bb x$.

\section{Problem Setup} \label{section:problem-setup}
\subsection{System Model}

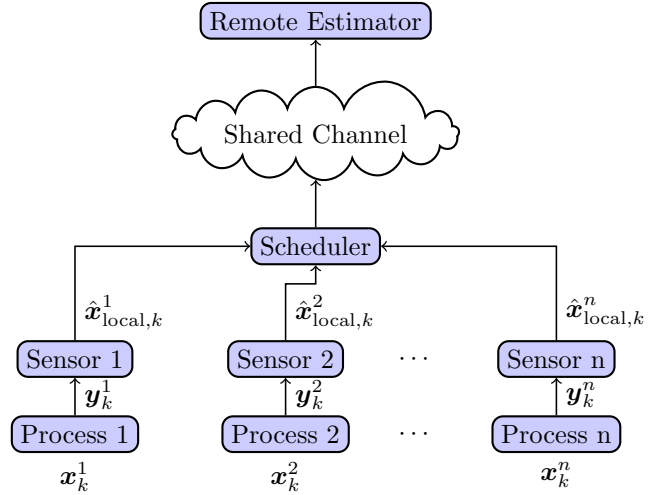
\begin{figure}
  \centering
  \begin{tikzpicture}[block/.style={rectangle,rounded corners,draw=black,fill=blue!20,thick}]

    \node[block,label=below:$\bb x_k^1$] (processa) at (0.2,-5) {Process 1};
    \node[block] (sensora) at (0.2,-4) {Sensor 1};
     edge [<-,semithick] (processa);
    \draw [semithick,<-] (sensora)--node[midway,right]{$\bb y_k^{1}$}(0.2,-4.7)--(processa);

    \node[block,label=below:$\bb x_k^2$] (processb) at (3.0,-5) {Process 2};
    \node[block] (sensorb) at (3.0,-4) {Sensor 2};
    \draw [semithick,<-] (sensorb)--node[midway,right]{$\bb y_k^{2}$}(3,-4.7)--(processb);

    \node[text width=2cm] at (5.5,-4) {$\cdots$};
    \node[text width=2cm] at (5.5,-5) {$\cdots$};

    \node[block,label=below:$\bb x_k^n$] (processn) at (6.6,-5) {Process n};
    \node[block] (sensorn) at (6.6,-4) {Sensor n};
    \draw [semithick,<-] (sensorn)--node[midway,right]{$\bb y_k^{n}$}(6.6,-4.7)--(processn);

    \node[block] (scheduler) at (3.4,-2.5) {Scheduler};

    \draw [semithick,->] (sensora)--node[midway,right]{$\hat{\bb x}_{{\rm local},k}^{1}$}(0.2,-3)--(0.2,-2.5)--(scheduler);
    \draw [semithick,->] (sensorb)--node[midway,right]{$\hat{\bb x}_{{\rm local},k}^{2}$}(3,-3)--(3.4,-3)--(scheduler);
    \draw [semithick,->] (sensorn)--node[midway,right]{$\hat{\bb x}_{{\rm local},k}^{n}$}(6.6,-3)--(6.6,-2.5)--(scheduler);

    \node[cloud, thick, cloud puffs=15.7, cloud ignores aspect, minimum width=3cm, minimum height=1cm, align=center, draw] (network) at (3.4,-1) {Shared Channel};

    \draw [semithick,<-] (network)--(scheduler);
    \node[block] (scheduler) at (3.4,0.5) {Remote Estimator} edge [<-,semithick] (network);

    \end{tikzpicture}
    \caption{System Block Diagram.}
\label{fig:system-frame-work-multi-sensor}
    \end{figure}

Denote $\mathcal Q:=\{1,\ldots,n\}$ as the index set of the processes or sensors. Consider the following $n$ linear time-invariant (LTI) systems (see Fig.~\ref{fig:system-frame-work-multi-sensor}):
\begin{subequations}
\begin{align}
\bb x_{k+1}^{i} & =  {\bb A}_i\bb x_k^i + \bb w_k^i, \label{eqn:system-dynamics} \\
\bb y_{k}^i & = {\bb C}_i \bb x_k^i + \bb v_k^i, \label{eqn:sensor-dynamics}
\end{align}
\end{subequations}
where $i\in\mathcal Q$, $\bb x_k^i \in \mathbb{R}^{n_i}$ is the state of the $i$th process at time $k$, $\bb y_{k}^i \in \mathbb{R}^{m_i}$ is the measurement obtained by the $i$th sensor at time $k$. The system noise $\bb w_k^i$'s, the measurement noise $\bb v_k^i$'s and the initial system state $\bb x_0^i$ are mutually independent zero-mean Gaussian random variables with covariances ${\bb Q}_i \geq \bb 0$, ${\bb R}_i > \bb 0$, and ${\bb \Pi}_i\geq \bb 0$, respectively. Assume that $({\bb A}_i, \sqrt{\bb Q}_i)$ is controllable and $({\bb A}_i, {\bb C}_i)$ is observable. Furthermore, we assume that ${\bb A}_i$'s are unstable since only unstable systems bring out stability issues rather than stable ones do and estimation performance will become unpredictable if left unattended too long.

Each sensor measures its corresponding system state and generates a local estimate first. More specifically, each time $k$ the $i$th sensor runs a Kalman filter to compute the minimum mean squared error (MMSE) estimate of $\bb x_k^i$. The information set of the $i$th sensor at time $k$ is given as:
  \begin{equation*}
    \mathcal{I}_{{\rm local},k}^i :=\{y_0^i, \ldots, y_k^i\},
  \end{equation*}
  with $\mathcal{I}_{{\rm local},-1}^i:=\emptyset$. Let us define
\begin{align*}
    \hat{\bb x}_{{\rm local},k}^{i-} &:= \mathbb{E}[{\bb x}_k^i| \mathcal{I}_{{\rm local},k-1}^i],\\
    \hat{\bb x}_{{\rm local},k}^i &:= \mathbb{E}[{\bb x}_k^i| \mathcal{I}_{{\rm local},k}^i],\\
{\bb P}_{{\rm local},k}^{i-} &:= \mathbb{E}\left[({\bb x}_k^{i}-\hat{\bb x}_{{\rm local},k}^{i-})({\bb x}_k^{i}-\hat{\bb x}_{{\rm local},k}^{i-})^{\top}| \mathcal{I}_{{\rm local},k-1}^i\right],\\
    {\bb P}_{{\rm local},k} &:= \mathbb{E}\left[({\bb x}_k^i-\hat{\bb x}_{{\rm local},k}^i)
    ({\bb x}_k^i-\hat{\bb x}_{{\rm local},k}^i)^{\top}|\mathcal{I}_{{\rm local},k}^i\right].
\end{align*}
Recall the standard Kalman filter \cite{anderson79}:
\begin{align*}
\hat{\bb x}_{{\rm local},k}^{i-} & =  {\bb A}_i \hat{\bb x}_{{\rm local},k-1}^{i}, \\
{\bb P}_{{\rm local},k}^{i-} & =  {\bb A}_i{\bb P}_{{\rm local},k-1}^{i}{\bb A}_i^{\top} + {\bb Q}_i, \\
{\bb K}_{{\rm local},k}^{i} & =  {\bb P}_{{\rm local},k}^{i-}{\bb C}_i^{\top}({\bb C}_i{\bb P}_{{\rm local},k}^{i-}{\bb C}_i^{\top} + {\bb R}_i)^{-1}, \\
\hat{\bb x}_{{\rm local},k}^{i} & =  \hat{\bb x}_{{\rm local},k}^{i-} + {\bb K}_{{\rm local},k}^{i}({\bb y}_k^{i} - {\bb C}_i\hat{\bb x}_{{\rm local},k}^{i-}), \\
{\bb P}_{{\rm local},k}^{i} & =  ({\bb I}_{n_i}-{\bb K}_{{\rm local},k}^{i}{\bb C}_i){\bb P}_{{\rm local},k}^{i-},
\end{align*} where the initial conditions are $\hat{\bb x}_{{\rm local},0}^{i-} = \bb 0$ and ${\bb P}_{{\rm local},0}^{i-} = {\bb \Pi}_i$. Then the sensors transmit their local estimate to the remote estimator. Due to the bandwidth limit, only \emph{one} sensor can access the shared channel, sending its estimate to the remote estimator, while the others have to discard their transmissions. The centralized scheduler makes a scheduling decision to determine which sensor shall be granted access to the channel.
Let $\theta:=\{\theta(k)\},~\theta(k):\mathbb{Z}\mapsto \mathcal{Q}$
be a \emph{time-based} sensor transmission schedule that specifies the scheduling decisions for all sensors according to time $k$.
Denote as $s_k(\theta)\in\mathcal Q$ the scheduling decision at time $k$ under a given $\theta$. If $s_k(\theta)=i$, only $\hat{\bb x}_{{\rm local},k}^i$ is transmitted to the estimator and $\hat{\bb x}_{{\rm local},k}^{j}$ is kept for $j\neq i$. Mathematically,
$$s_k(\theta)=\theta(k).$$
To study collision-free sensor scheduling protocols, {we restrict our attention to the set of time-based schedules.} Let $\Theta$ be the set of all feasible time-based
schedules.
The scheduling decision $s_k(\theta)$ under a given $\theta\in\Theta$ is abbreviated as $s_k$ unless otherwise noted.
Denote the indicator function $\mathbbm{1}_k^{i}(\theta)$ as the individual medium access indicator, \ie,
\[\mathbbm{1}_k^{i}(\theta)= \left\{
  \begin{array}{ll}
    1, & \hbox{if } s_k(\theta)=i;\\
    0, & \hbox{otherwise.}
  \end{array}
\right.
\]
Then we define a collection of the information sets $\mathcal I_k^{i}(\theta)$'s at the estimator as
\[\mathcal I_k^{i}(\theta) :=
\{\mathbbm{1}_0^{i}(\theta)\hat{\bb x}_{{\rm local}, 0}^i, \ldots, \mathbbm{1}_k^{i}(\theta)\hat{\bb x}_{{\rm local},k}^i\},~i\in\mathcal Q.\]
Due to the mutual independency among the systems and the fact that
$\mathbbm{1}_k^{i}(\theta)$ is a function of $k$,
conditioned on $\mathcal I_k^i(\theta)$, the estimator computes $\hat{\bb x}_k^i(\theta):=\mathbb{E}[{\bb x}_k^{i}|\mathcal I_k^i(\theta)]$, the MMSE estimate of the state ${\bb x}_k^i$ as follows:
\begin{align*}
\hat{\bb x}_k^{i}(\theta) =\mathbbm{1}_k^{i}(\theta)\hat{\bb x}_{{\rm local},k}^{i} + (1-\mathbbm{1}_k^{i}(\theta)){\bb A}_i\hat{\bb x}_{k-1}^{i}(\theta).
\end{align*}
In other words, the remote estimator updates the estimate according to $\hat{\bb x}_{{\rm local},k}^{i}$ if the packet sent by the $i$th sensor arrives; or simply runs a prediction step otherwise. The corresponding
estimation error covariance ${\bb P}_k^i(\theta):=  \mathbb{E}[({\bb x}_k^{i}-\hat{\bb x}_{k}^i)({\bb x}_k^{i}-\hat{\bb x}_{k}^i)^{\top}|\mathcal I_k^i(\theta)]$ is computed as
\begin{multline*}
{\bb P}_k^{i}(\theta) = \mathbbm{1}_k^{i}(\theta) {\bb P}_{{\rm local},k}^{i}\\
+(1-\mathbbm{1}_k^{i}(\theta))( {\bb A}_i{\bb P}_{k-1}^{i}(\theta){\bb A}_i^{\top} + {\bb Q}_i).
\end{multline*}

In this work we use the overall averaged estimation error covariance as a performance metric. For a given schedule $\theta$, define
$$W(\theta, T)=\sum_{k=0}^{T-1}\sum_{j=1}^{\card{\mathcal Q}} \left(\tr{{\bb P}_k^j(\theta)}\right),$$
and define the cost function $J(\theta)$ over an infinite horizon as
\begin{equation}\label{eqn:cost-function-definition}
J(\theta) := \limsup_{T\rightarrow\infty} \frac{1}{T}W(\theta, T).
\end{equation}
We are interested in searching a schedule $\theta\in\Theta$ which minimizes
 the sum of the trace of the averaged estimation error covariance of each system subject to the \emph{single-access} bandwidth constraint. Equivalently, we turn to solve the following problem:
\begin{problem} \label{problem:main-problem-1}
\begin{subequations}
\begin{align}
\min_{\theta \in \Theta}& ~~~~J(\theta) \label{eqn:7}\\
\mathrm{s.t.}&~~~\sum_{i=1}^{\card{\mathcal Q}}\mathbbm{1}_k^i(\theta) = 1,~ \forall k\in \hlb{\mathbb Z_+}.\label{eqn:8}
\end{align}
\end{subequations}
\end{problem}

Notice that when the constraint~\eqref{eqn:8} is relaxed into $\sum_{i=1}^{\card{\mathcal Q}}\mathbbm{1}_k^i(\theta)  \leq 1$, the optimal solution remains the same since any transmission is always preferred. A schedule $\theta^\ast \in \Theta$ is said to be optimal to Problem~\ref{problem:main-problem-1} if for any other $\theta\in \Theta$, $J(\theta^\ast) \leq J(\theta)$. In the sequel, we will abbreviate $\hat{\bb x}_k^i(\theta)$ as $\hat{\bb x}_k^i$ and ${\bb P}_k^i(\theta)$ as ${\bb P}_k^i$, etc., when the underlying schedule $\theta$ is evident from the context.

\section{Optimal Sensor Schedule}\label{section:optimal}
Before proceeding to solve the optimal scheduling problem, we first look into how the estimation error evolves in the bandwidth-limited case. For simplicity, we define the functions $h_i$ and $g_i$: $\mathbb{S}_{+}^{n_i}\mapsto \mathbb{S}_{+}^{n_i}$ as follows:
\begin{align*}
h_i(\bb X) & :=  {\bb A}_i{\bb X}{\bb A}_i^{\top} + {\bb Q}_i, \\
g_i(\bb X) & :=  {\bb X} - {\bb X}{\bb C}_i^{\top}({\bb C}_i{\bb X}{\bb C}_i^{\top} + {\bb R}_i)^{-1}{\bb C}_i{\bb X}.
\end{align*}
Then the following recursive equation can be used to determine ${\bb P}_{{\rm local},k}^i$ at the $i$th sensor at time $k$,
\[
{\bb P}_{{\rm local},k}^{i} = g_i\circ h_i({\bb P}_{{\rm local},k-1}^{i}).
\]
With the assumptions that $({\bb A}_i, \sqrt{\bb Q}_i)$ is controllable and $({\bb A}_i, {\bb C}_i)$ is observable, there exists a unique solution ${\bb P}_i \geq \bb 0$ to the following discrete algebraic Riccati equation \cite{anderson79}:
\begin{equation}\label{eqn:steady-state-error-covariance}
{\bb P}_i = g_i\circ h_i({\bb P}_i),
\end{equation}
which is referred to as the steady-state estimation error covariance of the Kalman filter at the $i$th sensor. Since ${\bb P}_{{\rm local},k}^{i}$ converges to ${\bb P}_i$ exponentially fast, we assume the initial state covariance ${\bb \Pi}_i = {\bb P}_i$ without loss of generality. As a result ${\bb P}_{{\rm local},k}^{i} = {\bb P}_i,~\forall k\in\mathbb{Z}$. Consequently, the estimation error covariance ${\bb P}_k^i$ at the remote estimator satisfies
\begin{align}
{\bb P}_k^{i}= \mathbbm{1}_k^i(\theta) {\bb P}_i+(1-\mathbbm{1}_k^i(\theta)) h_i({\bb P}_{k-1}^{i}).\label{eqn:covariance_update}
\end{align}
Notice that the estimation error covariance ${\bb P}_k^{i}$ grows exponentially fast if no estimate $\hat{\bb x}_k^{i}$ arrives at the estimator. As only a single sensor can access the medium at the same time, the conflict between the traffic control and the estimation quality must be resolved in the way of average optimum.
For the above definitions, we have the following lemma.
\begin{lemma}[ {\cite[Lemma~3.1]{shi2012scheduling}}]\label{lemma:supporting-lemma}
The following statements hold for any $i\in\mathcal Q$:
\begin{enumerate}
\item[(i).] For any $\ell_1,\ell_2\in\mathbb{Z}_+$ with $\ell_1 \leq \ell_2$, $h_i^{\ell_1}({\bb P}_i) \leq h_i^{\ell_2}({\bb P}_i)$
\item[(ii).] For any $\ell \in \mathbb{Z}_+$,
\[\tr{{\bb P}_i} < \tr{h_i({\bb P}_i)}< \cdots < \tr{h_i^{\ell}({\bb P}_i)}.\]
\end{enumerate}
\end{lemma}

\subsection{Boundedness of Off-duty Duration}

The direct construction of an optimal schedule is difficult. Instead we first derive an upper bound of the maximum off-duty period of each sensor for an optimal schedule. By significantly reducing the size of the set of feasible solutions, we formulate an average-cost Markov decision process problem to search the optimal schedule.

%
%

From \eqref{eqn:covariance_update} we notice that the time length between two consecutive transmissions of the same sensor affects the estimation error as the error covariance will be reset once a transmission is scheduled. Hence it is worthwhile to study the relationship between the length of off-duty duration and the optimality of a scheduler.
Mathematically, the off-duty duration of the $i$th sensor is denoted as follows.  For the scheduler $\theta$, define a sequence of times, the time
instants at which the $i$th sensor is scheduled to transmit, as
\begin{align*}
\tau_1^i&:=\min\{k: k\in \mathbb{Z}, s_k(\theta)=i\},\\
\tau_2^i&:=\min\{k:k>\tau_1^i,s_k(\theta)=i\},\\
&~~\vdots\\
\tau_j^i&:=\min\{k:k>\tau_{j-1}^i,s_{k_j}(\theta)=i\}.
\end{align*}
\hlb{Consider the time horizon $[1,T]$ and let $$\mathcal S_i(\theta,T):=\{\tau^i_1,\ldots,\tau^i_{\sigma_i(T)}\}$$
represent the set of all transmissions of the $i$th sensor over $[1,T]$ given $\theta$, where $\sigma_i(T)$ denotes the total number of transmissions of the $i$th sensor by time $T$. Note that $|\mathcal S_i(\theta,T)|=\sigma_i(T)$ and $\tau^i_{\sigma_i(T)}\leq T$ while $\tau^i_{\sigma_i(T)+1}>T$.}

Define the off-duty duration between two consecutive transmissions over $[1,T]$ as
\begin{align}
d_1^i(\theta,T)&=T-\tau^i_{\sigma_i(T)}+\tau^i_1,\notag\\
d_2^i(\theta,T)&=\tau_2^i-\tau_1^i,\notag\\
&~~\vdots\notag\\
d_{\sigma_i(T)}^i(\theta,T)&=\tau^i_{\sigma_i(T)}-\tau^i_{\sigma_i(T)-1}.
\end{align}
\begin{remark}
\hlb{Note that $d_1^i(\theta,T)$ is the $T$'s complement of $\tau^i_{\sigma_i(T)}-\tau^i_1$. This definition makes more sense in the context of periodic schedules with period of $T$, \ie, the off-duty duration between the last transmission in the $j$th period and the first transmission in the $(j+1)$th period.}
\end{remark}
By stacking all $d_j^i(\theta,T)$'s of the $i$th sensor into a vector, we have $\bb{d}^i(\theta,T)=\left[d_1^i(\theta,T),\ldots,d_{\sigma_i(T)}^i
(\theta,T)\right]^{\top}$. For example, considering a schedule $\theta$ for a four-sensor case over a horizon of length $8$:
\begin{align*}
&s_{1}(\theta)=3,s_{2}(\theta)=1,s_{3}(\theta)=4,s_{4}(\theta)=3,\\
&s_{5}(\theta)=1,s_{6}(\theta)=2,s_{7}(\theta)=3,s_{8}(\theta)=1,
\end{align*}
we have $\mathcal S_1(\theta,8):=\{2,5,8\}$ and $\bb{d}^1(\theta,T)=\left[2~3~3\right]^{\top}$.

We now give a necessary condition for optimality.
\begin{theorem}\label{theorem:maximum_gap}
Let $\mathcal M_i(T):=\{\tau_\ell^i\in\mathcal S_i(\theta^*,T): d_\ell^i(\theta^*,T)>\Delta_{i}^{\max}\}$,
 where $\Delta_{i}^{\max}$ is given by
\begin{align}
\Delta_{i}^{{\rm max}}&=\max \{ \Delta_{j,i}^{{\rm max}}:j\in\card{\mathcal Q},j\neq i\},\label{eqn:2}\\
\Delta_{j,i}^{{\rm max}}& =\max\big\{3\card{\mathcal Q}-2,\max\{\ell_1+\ell_2+\ell_3+1:\nonumber\\
&~~~~~~~~~~~\tr{\sum_{l=0}^{\ell_3-1}h_i^\ell(h_i^{\ell_1+\ell_2}({\bb P}_i)-{\bb P}_i)} \nonumber\\
&~~~~~~~~~~~~~~~\leq\tr{\sum_{\ell=0}^{\ell_3-1}h_j^\ell(h_j^{\ell_2}({\bb P}_j)-{\bb P}_j)}\left.\right\}\left.\right\}\nonumber
\end{align}
with $\ell_1\in \hlb{\mathbb Z_{++}},~\ell_2,\ell_3\in\{1,\ldots,3|\mathcal{Q}|-4\}.$
For any optimal schedule $\theta^*\in \Theta$,  $\lim_{T\rightarrow\infty}\card{\mathcal M_i(T)}/T=0,~\forall i\in\mathcal Q$.
\end{theorem}
\begin{pf*}{Proof.}
\hlb{We shall prove by contradiction. In other words, by violating the above results we can construct a schedule such that its cost function is smaller than the optimal one.}

\hlb{Consider the schedule $\theta$ and suppose the $i$th sensor violates the necessary condition,
\ie, $$\limsup_{T\rightarrow\infty}\frac{\card{\mathcal M_i(T)}}{T}>0.$$
Without loss of generality, let $\tau_L^i\in \mathcal M_i(T)$, which means
$$d_L^i(\theta,T)\geq \Delta_i^{\max}\geq 3|\mathcal{Q}|-2.$$
Take an interval $[\tau_{L}^i-3|\mathcal{Q}|+3, \tau_{L}^i-1]\subset[\tau_{L-1}^i,\tau_{L}^i]$.
There must exist one sensor out of the remaining $\card{\mathcal Q}-1$ sensors which is scheduled for at least three transmissions within $[\tau_{L}^i-3|\mathcal{Q}|+3, \tau_{L}^i-1]$. Without loss of generality, let the $j$th sensor transmits at least three times and the last three transmissions to be $s_{\tau_{L-1}^i+v_1}=s_{\tau_{L-1}^i+v_2}=s_{\tau_{L-1}^i+v_3}=j$, where $d_{L}^i(\theta,T)-3|\mathcal{Q}|+3\leq v_1<v_2<v_3\leq d_{L}^i(\theta,T)-1$. For better understanding, we illustrate the notations on the time axis in Fig. \ref{fig:thm1}.}

\begin{figure}
  \centering
  \begin{tikzpicture}
  \draw [->,thick] (0,0) -- (7,0);
  \node at (6.8,0.2) {$\cdots$};

  \draw [->,color=blue,thick](0.21,0) -- (0.21,0.4); \node at (0.21,-0.3) {$\tau_{L-1}^i$};

  \draw [->,color=blue,thick] (6.2,0) -- (6.2,0.4); \node at (6.2,-0.3) {$\tau_{L}^i$};
  \draw [->,color=red,thick] (1.5,0) -- (1.5,0.4); \node at (1.5,-0.2) {$\tau_{L-1}^i+v_1$};
  \draw [->,color=red,thick] (3.3,0) -- (3.3,0.4);
  \node at (3.3,-0.3) {$\tau_{L-1}^i+v_2$};
  \draw [->,color=red,thick] (5,0) -- (5,0.4);
  \node at (5,-0.2) {$\tau_{L-1}^i+v_3$};

  \draw [dashed,<-,color=black] (0.3,0.3) -- (0.6,0.3);\node at (0.8,0.3) {$\ell_1$};\draw [dashed,->,color=black] (1,0.3) -- (1.3,0.3);

  \draw [dashed,<-,color=black] (1.8,0.3) -- (2.2,0.3);\node at (2.4,0.3) {$\ell_2$};\draw [dashed,->,color=black] (2.6,0.3) -- (2.9,0.3);

  \draw [dashed,<-,color=black] (3.7,0.3) -- (4.0,0.3);\node at (4.2,0.3) {$\ell_3$};\draw [dashed,->,color=black] (4.4,0.3) -- (4.7,0.3);
  \end{tikzpicture}
  \caption{Illustration of notations on the time axis.}
  \label{fig:thm1}
\end{figure}
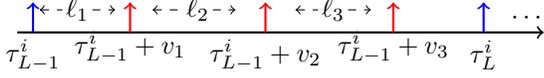

By constructing a schedule $\theta'\in\Theta$ to be the same with $\theta$ except that $s_{\tau_{L-1}^i+v_2}=i$, for any $T\geq \tau_{L}^i,$
\begin{align}
&W(\theta,T)-W(\theta',T)\notag\\
=&\sum_{k=0}^{T-1}\sum_{\ell=1}^{\card{\mathcal Q}} \left(\tr{{\bb P}_k^\ell(\theta)}-\tr{{\bb P}_k^\ell(\theta')}\right)\nonumber\\
=&\mathrm{Tr}\left[\sum_{l=0}^{d_{L}^i(\theta,T) -v_2-1}h_i^\ell(h_i^{v_2}({\bb P}_i)-{\bb P}_i)\right.\nonumber\\
&~~~~~~~~~~~~~~~~~~~-\left.\sum_{\ell=0}^{v_3-v_2-1}h_j^\ell
(h_j^{v_2-v_1}({\bb P}_j)-{\bb P}_j)\right]\nonumber\\
> & \mathrm{Tr}\left[\sum_{\ell=0}^{v_3-v_2-1}
h_i^\ell(h_i^{v_2-v_1+v_1}({\bb P}_i)-{\bb P}_i)\right.\nonumber\\
&~~~~~~~~~~~~~~~~~~~-\left.\sum_{\ell=0}^
{v_3-v_2-1}h_j^\ell(h_j^{v_2-v_1}({\bb P}_j)-{\bb P}_j)\right].\label{eqn:14}
\end{align}
\hlb{Therefore, if
\begin{align}v_3>&\max\big\{\ell_1+\ell_2+\ell_3:\tr{\sum_{\ell=0}^{\ell_3-1}h_i^\ell(h_i^{\ell_1+\ell_2}({\bb P}_i)-{\bb P}_i)} \nonumber\\
&~~~~~~~~~~~~~~~~\leq\tr{\sum_{\ell=0}^{\ell_3-1}h_j^l(h_j^{\ell_2}({\bb P}_j)-{\bb P}_j)}\big\}\label{eqn:20}
\end{align}
with $\ell_1\in {\mathbb Z_{++}},~\ell_2,\ell_3\in\{1,\ldots,3|\mathcal{Q}|-4\}$,
or equivalently, if $d_L^i(\theta,T)\geq
\Delta_{j,i}^{\max}$, then $W(\theta,T)-W(\theta',T)\geq c>0$, where $c$ is given by
\begin{align*}
&\min\Bigg\{\tr{\sum_{\ell=0}^{\ell_3-1}h_i^\ell(h_i^{\ell_1+\ell_2}({\bb P}_i)\hspace{-1mm}-\hspace{-1mm}{\bb P}_i)-\hspace{-0.5mm}\sum_{l=0}^{\ell_3-1}h_j^\ell(h_j^{\ell_2}({\bb P}_j)\hspace{-1mm}-\hspace{-1mm}{\bb P}_j)}:\\
& \sum_{j=1}^3\ell_j\geq \Delta_{i}^{\max},\ell_1\in \hlb{\mathbb Z_{++}},~\ell_2,\ell_3\in\{1,\ldots,3|\mathcal{Q}|-4\}\Bigg\}:=c.
\end{align*}
Note that $\ell_1$ in \eqref{eqn:20} must be bounded since $A_i$ is unstable and $\ell_2,\ell_3$ are bounded.}

Taking all elements in~$\mathcal M_i(T)$ into account and
constructing a schedule $\theta_{*}$ following a similar
procedure to above, we obtain
$$W(\theta,T)-W(\theta_*,T)\geq c |\mathcal M_i(T)|.$$
As $T\to \infty$,
\begin{align*}J(\theta)-J(\theta_*)=&\limsup\frac{W(\theta,T)-W(\theta_*,T)}{T}\\
\geq & \limsup \frac{c |\mathcal M_i(T)|}{T}>0,
\end{align*}
which implies that the schedule $\theta$ violates the optimality and thus completes the proof.
\end{pf*}

{
Thanks to Theorem~\ref{theorem:maximum_gap}, we are able to restrict the length of the off-duty durations for all sensors to be bounded throughout the infinite horizon without loss of performance.
It helps curtail the space of feasible schedules, over which an optimal schedule is searched. This observation is a basis of designing algorithms to search  optimal or suboptimal schedules.
It is interesting to find a tighter bound for $d_k^i(\theta^*,T)$, which is left as a further work.}

\subsection{Markov Decision Process Description}
{In this part we will model a Markov decision process (MDP) problem to calculate the optimal schedule $\theta^\ast\in\Theta$. Denote as
\begin{align*}
\mathcal V:=\{\nu = [v_1,  \ldots,  v_{\card{\mathcal Q}}]^{\top}:~v_i=&1,\ldots, \Delta_i^{\rm max};\\
&v_i\neq v_j,\forall i\neq j;~\exists v_i=1\}
\end{align*}
the MDP state space and each $v_i$ represents
the time between the current time and the most recent
instance when the $i$th sensor is scheduled to transmit,
\ie, $v_i=\inf \{k-\ell:s_\ell=i,\ell<k\}$ at time $k$. The decision maker chooses an action $\alpha_k$ at time $k$ from the set of actions $\mathcal A:=\{a_1,\ldots,a_{\card{\mathcal Q}}\}$
and $\alpha_k=a_j$ means that the $j$th sensor is scheduled to transmit at time $k$. If the action $\alpha=a_i$ for state $\nu$ is assumed, then the next state $\nu'$ will satisfy $\nu'[i]=1$ and $\nu'[j]=\nu[j]+1$ for $j\neq i$. Therefore, we have the set of \emph{allowable} actions for state $\nu$ is
\begin{align*}
\mathcal A_{\nu}:= \{a_1,\ldots,a_{\card{\mathcal Q}}\}\backslash \overline{\mathcal A}_{\nu},
\end{align*}
where $\overline{\mathcal A}_{\nu}:=\{a_i\in\mathcal A: \nu'\notin \mathcal V, \hbox{~where~}\nu'[i]=1,~\nu'[j]=\nu[j]+1,\forall j\neq i\}$ ensuring that the possible next states from $\nu$ must remain within the finite state space.
The transition probability from state $\nu$ to state $\nu'$ under action $\alpha=a_i$ is defined as
\begin{multline*}
\Pr(\nu'|\nu,a_i)\\
=
\left\{
  \begin{array}{ll}
    1, & \hbox{if $\nu'[i]=1$ and $\nu'[j]=\nu[j]+1$ for $j\neq i$;} \\
    0, & \hbox{otherwise.}
  \end{array}
\right.
\end{multline*}
The reward $r:\mathcal V\mapsto \mathbb R$, independent of the action, is defined as
\[
r(\nu)=-\tr{\sum_{i=1}^{\card{\mathcal Q}}h_i^{v_i-1}({\bb P}_i)}, \hbox{~for~}
\nu=[v_1,\ldots,v_{\card{\mathcal Q}}]^{\top}.
\]
}
A decision at time $k$ is a mapping $u_k:\mathcal V\mapsto \mathcal A_{\nu}$ and the set of admissible decisions is denoted as $\mathcal U$. We  model the problem in \eqref{eqn:cost-function-definition} as an average-cost MDP problem $\mathscr M:=\left(\mathcal V,\mathcal A_{\nu},\Pr(\cdot|\cdot,\cdot),r(\cdot)\right)$, where $\mathcal V$ and $\mathcal A_{\nu}$ are both finite sets.
A policy $\mu$ for $\mathscr M$ is a sequence of decision rules
$(u_1,u_2,\ldots)$. The policy $\mu$ is said to be \emph{stationary}
if $u_k=u$ for all $k\in\mathbb{Z}$.
We thus define the \emph{average expected reward} of the policy $\mu$ by
\begin{align}
g_{\mu}(\nu_0)=\lim_{T\rightarrow \infty}\frac{1}{T}\mathbb E_{\nu_0}^{\mu}\left[\sum_{k=0}^{T-1}r(V_k)\right],\label{eqn:modified_cost}
\end{align}
where $\nu_0$ is the initial state and $V_k$ is the state at time $k$, and the expectation is taken with respect to the policy $\mu$.
The target is to search a policy such that the average reward is
maximized. The policy $\mu^*$ is optimal for $\mathscr{M}$ if
\[g_{\mu^*}(\nu_0)\geq g_{\mu}(\nu_0),~\forall \nu_0\in\mathcal V.\]
\hlb{Notice that the MDP model is \emph{communicating}, namely, there exist policies under which each state is accessible from each other state. Then we can conclude that there must exist a stationary optimal policy with constant average reward based on \cite[Theorem 8.3.2, Theorem 9.1.8]{puterman2005markov}, \ie, $g_{\mu^*}(\nu)=g_{\mu^*}(\nu')$ for any $\nu,\nu'\in\mathcal V$. For simplicity, we denote the optimal average reward to be $g^*$ with a little abuse of notation. The following optimality equations are required to obtain the optimal policies \cite[Chapter 9]{puterman2005markov},
\begin{align}
&\sup_{\alpha\in\mathcal A_{\nu}}\left\{r(\nu)-g^*+\sum_{\nu'\in\mathcal V} \Pr(\nu'|\nu,\alpha)h(\nu')-h(\nu)\right\}\nonumber\\
&~~~~~~~~~~~~~~~~~~~~~~~~~~~~~~~~~~~~~~~~~~~~~~~~~~~~~~~~=0\label{eqn:optimality_2},
\end{align}
Then the following lemma states the existence of optimal policy and how to identify an optimal policy.
\begin{lemma}\label{theorem:optimal_policy}
The following statements hold:
\begin{enumerate}
  \item[(i).] There always exists a solution to $g^*,h\in \mathcal W$ satisfying  \eqref{eqn:optimality_2}.
  \item[(ii).] \label{theorem:policy_c}Suppose $g,h\in \mathcal W$ satisfy  \eqref{eqn:optimality_2}. If \[u^*(\nu)\in \arg\max_{\mu}\{r(\nu)+\sum_{\nu'\in\mathcal V} \Pr(\nu'|\nu,\alpha)h(\nu')\},\] then $\mu^*:=(u^\ast,u^\ast,\ldots)$ is optimal
     for $\mathscr{M}$.
\end{enumerate}
\end{lemma}
\begin{pf*}{Proof.}
As $\mathcal V$ and $\mathcal A$ are finite, we directly derive $(i)$ from \cite[Theorem 9.1.4]{puterman2005markov}, which states the existence of the solution to  \eqref{eqn:optimality_2}. By knowing the optimal solutions to $g^*$ and $h$, the statement $(ii)$ shows how to identify the optimal policies based on \cite[Theorem 9.1.7]{puterman2005markov}.
\end{pf*}}
The optimal policy $\mu^\ast$ can be searched by linear programming, value iteration or policy iteration \cite{puterman2005markov}.

\hlb{To see the equivalence between Problem \ref{problem:main-problem-1} and the MDP problem, let the scheduler in the original problem be the decision maker in the MDP. At each time the scheduler identifies which state it is in and takes the optimal action to transit into the next state. Due to the equivalence between minimizing the cost function in \eqref{eqn:cost-function-definition} and maximizing the reward function in \eqref{eqn:modified_cost}, we can conclude that $J(\theta^*)=g^*$.}

The next theorem shows that there exists an optimal schedule converging to be periodic and the schedule can be easily determined offline. The result relies on the finiteness of the state space $\mathcal V$.

\begin{theorem}[Asymptotic periodicity]\label{theorem:periodic}
There always exists an optimal schedule $\mu^*$ converging to be periodic, i.e., for the scheduler $\theta^\ast$ generated by $\mu^*$, $\exists L,M\in\mathbb{Z},$ such that $s_k(\theta^\ast)=
      s_{k+L}(\theta^\ast)$ for all $k\geq M$.
\end{theorem}
\begin{pf*}{Proof.}
If the number of optimal actions exceeds $1$ for some $v\in\mathcal V$ under $\mu^*$ in \ref{theorem:policy_c}) of Lemma \ref{theorem:optimal_policy}, one arbitrarily picks up one as the optimal action. Thus without loss of generality, assume that $\mu^*$ is deterministic, i.e., $\mu^*$ maps one state to a single action. Since $\mathcal V$ is finite, given a $\mu^*$ there must exist a recurrent state which the system will re-enter after a finite number of time instants. In other words, the system will repeatedly follow a trajectory of states.
Furthermore, the period is less than the cardinality of the state space $\mathcal V$.
\end{pf*}

Owing to Theorem \ref{theorem:periodic}, without loss of performance, we can restrict our attention to the class of \emph{periodic schedules} $$\Theta_{\mathscr P}:=\{\theta\in\Theta:\exists L \in\mathbb{Z}\hbox{~s.t.~} s_k(\theta)=s_{k+L}(\theta),~\forall k\in\mathbb{Z}\}.$$
Moreover, for any $\theta\in \Theta_{\mathscr P}$,
$\lim_{T\to\infty}W(\theta,T)/T$ exists and certainly
$$J(\theta)=\lim_{T\to\infty}\frac{1}{T}W(\theta,T).$$
\subsection{Uniformity of Optimal Sensor Scheduling}
On the basis of the periodicity of optimal schedules established in Theorem~\ref{theorem:periodic},
we will present  a structural description for the optimal sensor
schedule.
In fact, any schedule that violates the law of \emph{uniformity} must not be optimal.
This result is  elaborated using a vector majorization argument.
Before we present the result, the definition of vector majorization is first introduced.
\begin{definition}[Vector majorization]\label{Def:majorization}
Denote $\bb a,\bb b\in {\mathbb R}^m$. Then
$\bb a$ is said to be majorized by $\bb b$, denoted as $\bb a\prec \bb b$, if
\begin{align}
&\sum_{i=1}^j\bb{a}^{\downarrow}[i]\leq\sum_{i=1}^j\bb{b}^{\downarrow}[i]
\end{align}
holds for all $j=1,\ldots,m$ with
equality for $j=m,$
where $\bb{a}^{\downarrow}$ (or $\bb{b}^{\downarrow}$) is a vector
that has the same entries with $\bb{a}$ (or $\bb{b}$) but in a nonincreasing order.
\end{definition}
We are now ready to show the necessary uniformity condition for the optimal schedule.
\begin{theorem}[Uniformity]\label{theorem:uniform_transmission}
Consider a schedule $\theta\in\Theta_{\mathscr P}$ with a period $L$. If there exists a schedule $\theta'\in\Theta_{\mathscr P}$ such that
\begin{enumerate}
\item[(i).]  $|\mathcal{S}_i(\theta,mL)|=|\mathcal{S}_i(\theta',mL)|~\forall i\in\mathcal Q$, and
\item[(ii).] $\bb{d}^i(\theta',mL)\prec \bb{d}^i(\theta,mL)~\forall i\in\mathcal Q$, and
 \item[(iii).] $\exists j\in\mathcal Q, \bb{d}^j(\theta',mL)^{\downarrow}\neq \bb{d}^j(\theta,mL)^{\downarrow}$
\end{enumerate}
hold for some $m\in\mathbb{Z}_+$, then $J(\theta')<J(\theta)$
and $\theta$ is not an optimal solution to Problem~\ref{problem:main-problem-1}.
\end{theorem}
\begin{pf*}{Proof.}
Suppose there exists $\theta'$ such that the conditions $(i)$-$(iii)$ hold for some $m\in\mathbb{Z}_+$.
Then the conditions $(i)$-$(iii)$ hold for $m=1$ and vice versa.
Therefore, without loss of generality, let $m=1$.
Note that $\sigma_i(T)=\card{\mathcal S_i(\theta',L)}=\card{\mathcal S_i(\theta,L)}~\forall i\in\mathcal Q$. Fixing $i\in\mathcal Q$, we will show that
\begin{align}
&\lim_{T\to \infty}\frac{1}{T}\sum_{k=0}^{T-1} \tr{{\bb P}_k^i(\theta')}-
\lim_{T\to \infty}\frac{1}{T}\sum_{k=0}^{T-1} \tr{{\bb P}_k^i(\theta^*)} \nonumber\\
=& \frac{1}{L}\sum_{l=1}^{\sigma_i(T)}\sum_{j=1}^
{\bb{d}^i(\theta',L)^{\downarrow}[\ell]}\tr{h_i^{j-1}({\bb P}_i)} \nonumber\\
&~~~~~~~~~~~~~~~~~-\frac{1}{L}\sum_{l=1}^{\sigma_i(T)}
\sum_{j=1}^{\bb{d}^i(\theta,L)^{\downarrow}[\ell]}\tr{h_i^{j-1}({\bb P}_i)}\nonumber\\
\leq &0.\label{eqn:11}
\end{align}
First it is straightforward to see that
\begin{align*}
&\sum_{j=1}^{\bb{d}^i(\theta',L)^{\downarrow}[1]}\tr{h_i^{j-1}({\bb P}_i)}-\sum_{j=1}^{\bb{d}^i(\theta,L)^{\downarrow}[1]}\tr{h_i^{j-1}({\bb P}_i)}\\
=& -\sum_{j=\bb{d}^i(\theta',L)^{\downarrow}[1]+1}^{\bb{d}^i(\theta,L)^{\downarrow}[1]}\tr{h_i^{j-1}({\bb P}_i)}\\
\leq & -\sum_{j=\bb{d}^i(\theta',L)^{\downarrow}[2]+1}^
{\bb{d}^i(\theta',L)^{\downarrow}[2]+
d_1
}\tr{h_i^{j-1}({\bb P}_i)},
\end{align*}
where $d_1=\bb{d}^i(\theta,L)^{\downarrow}[1]-\bb{d}^i(\theta',L)^{\downarrow}[1]
$.
With the element-descending order of $\bb{d}^i(\theta',L)^{\downarrow}$, Definition~\ref{Def:majorization} and Lemma \ref{lemma:supporting-lemma}, we have
{
\begin{align}
&\sum_{l=1}^2\left(
\sum_{j=1}^{\bb{d}^i(\theta',L)^{\downarrow}[\ell]}\tr{h_i^{j-1}({\bb P}_i)}-
\sum_{j=1}^{\bb{d}^i(\theta,L)^{\downarrow}[\ell]}\tr{h_i^{j-1}({\bb P}_i)}\right)\notag\\
&\leq \sum_{j=1}^{\bb{d}^i(\theta',L)^{\downarrow}[2]}\tr{h_i^{j-1}({\bb P}_i)}-
\sum_{j=1}^{\bb{d}^i(\theta,L)^{\downarrow}[2]}\tr{h_i^{j-1}({\bb P}_i)}\notag\\
&~~~~~~~-\sum_{j=\bb{d}^i(\theta',L)^{\downarrow}[2]+1}^
{\bb{d}^i(\theta',L)^{\downarrow}[2]+
d_1
}\tr{h_i^{j-1}({\bb P}_i)}.\label{eqn:step2}
\end{align}
}
If $\bb{d}^i(\theta,L)^{\downarrow}[2]\leq \bb{d}^i(\theta',L)^{\downarrow}[2]$, we have that \eqref{eqn:step2} is not larger than
\begin{align*}
&\sum_{j=1}^{\bb{d}^i(\theta',L)^{\downarrow}[2]}\tr{h_i^{j-1}({\bb P}_i)}
-\sum_{j=1}^{\bb{d}^i(\theta,L)^{\downarrow}[2]+d_1}\tr{h_i^{j-1}
({\bb P}_i)}\\
&\leq - \sum_{j=\bb{d}^i(\theta',L)^{\downarrow}[2]+1}^{\bb{d}^i(\theta,L)^{\downarrow}[2]+d_1}\tr{h_i^{j-1}
({\bb P}_i)}\\
&\leq -
\sum_{j=\bb{d}^i(\theta',L)^{\downarrow}[3]+1}^
{\bb{d}^i(\theta',L)^{\downarrow}[3]+d_2+d_1}\tr{h_i^{j-1}
({\bb P}_i)},
\end{align*}
where $d_2=\bb{d}^i(\theta,L)^{\downarrow}[2]-\bb{d}^i(\theta',L)^{\downarrow}[2]$.
Otherwise, if $\bb{d}^i(\theta,L)^{\downarrow}[2]> \bb{d}^i(\theta',L)^{\downarrow}[2]$, then~\eqref{eqn:step2} is not larger than
\begin{align*}
-& \hspace{-3mm}\sum_{j=\bb{d}^i(\theta',L)^{\downarrow}[2]+1}^
{\bb{d}^i(\theta,L)^{\downarrow}[2]
}\hspace{-4mm}\tr{h_i^{j-1}({\bb P}_i)}
-\hspace{-3mm}\sum_{j=\bb{d}^i(\theta',L)^{\downarrow}[2]+1}^
{\bb{d}^i(\theta',L)^{\downarrow}[2]+
d_1
}\hspace{-4mm}\tr{h_i^{j-1}({\bb P}_i)}\\
\leq &-\hspace{-3mm}\sum_{j=\bb{d}^i(\theta',L)^{\downarrow}[3]+1}^
{\bb{d}^i(\theta',L)^{\downarrow}[3]+d_2
}\hspace{-4mm}\tr{h_i^{j-1}({\bb P}_i)}
-\hspace{-3mm}\sum_{j=\bb{d}^i(\theta',L)^{\downarrow}[3]+1}^
{\bb{d}^i(\theta',L)^{\downarrow}[3]+
d_1
}\hspace{-4mm}\tr{h_i^{j-1}({\bb P}_i)}.
\end{align*}
Thus by induction, eventually we obtain
\begin{align}
&\sum_{\ell=1}^{\sigma_i(T)}\left(\sum_{j=1}^{\bb{d}^i(\theta',L)
^{\downarrow}[\ell]}\tr{h_i^{j-1}({\bb P}_i)}\right.\notag\\
&~~~~~~~~~~~~~~~~~~~~~~~~~~~~ \left.-\sum_{j=1}^{\bb{d}^i(\theta,L)^{\downarrow}[\ell]}\tr{h_i^{j-1}({\bb P}_i)}\right)\notag\\
&\leq \sum_{j=1}^{\bb{d}^i(\theta',L)^{\downarrow}[\sigma_i
(T)]}\tr{h_i^{j-1}({\bb P}_i)}\notag\\
&~~~~
-\sum_{j=1}^{\bb{d}^i(\theta,L)^{\downarrow}[\sigma_i(T)]+d_1+\cdots+
d_{\sigma_i(T)-1}}\tr{h_i^{j-1}({\bb P}_i)},\label{eqn:stepn}
\end{align}
where $d_j=\bb{d}^i(\theta,L)^{\downarrow}[j]
-\bb{d}^i(\theta',L)^{\downarrow}[j]$.
Since $\sum_{j=1}^{\sigma_i(T)}d_j=0$, we conclude that
$\eqref{eqn:stepn}\leq 0$,
which implies the inequality in \eqref{eqn:11}.
Therefore, taking all the sensors into account, we have
$J(\theta')-J(\theta)<0$
with the strict inequality due to $(iii)$,
which completes the proof.
\end{pf*}

Theorem \ref{theorem:uniform_transmission} provides a necessary condition for optimality and thus can be used to identify an optimal schedule. The following example shows how it works.
\begin{example}
\begin{figure}
  \centering
  \begin{tikzpicture}
  \draw [->,thick] (0,0) -- (7,0);
  \node at (6.5,-0.3) {$s_k(\theta_1)$};
  \node at (0.5,0.2) {$\cdots$};
  \node at (5.8,0.2) {$\cdots$};
  \draw [dashed,->,color=blue,thick](1,0) -- (1,0.6);
  \draw [dashed,->,color=blue,thick] (1.8,0) -- (1.8,0.6);
  \draw [dashed,->,color=blue,thick] (2.6,0) -- (2.6,0.6);
  \draw [->,color=red,thick] (3.4,0) -- (3.4,0.4);
  \draw [->,color=red,thick] (4.2,0) -- (4.2,0.4);
  \draw [->,color=red,thick] (5,0) -- (5,0.4);

  \draw [->,thick] (0,-1) -- (7,-1);
  \node at (6.5,-1.3) {$s_k(\theta_2)$};
  \node at (0.5,-0.8) {$\cdots$};
  \node at (5.8,-0.8) {$\cdots$};
  \draw [dashed,->,color=blue,thick](1,-1) -- (1,-0.4);
  \draw [->,color=red,thick] (1.8,-1) -- (1.8,-0.6);
  \draw [->,color=red,thick] (2.6,-1) -- (2.6,-0.6);
  \draw [dashed,->,color=blue,thick] (3.4,-1) -- (3.4,-0.4);
  \draw [dashed,->,color=blue,thick] (4.2,-1) -- (4.2,-0.4);
  \draw [->,color=red,thick] (5,-1) -- (5,-0.6);

  \draw [->,thick] (0,-2) -- (7,-2);
  \node at (6.5,-2.3) {$s_k(\theta_3)$};
  \node at (0.5,-1.8) {$\cdots$};
  \node at (5.8,-1.8) {$\cdots$};
  \draw [dashed,->,color=blue,thick](1,-2) -- (1,-1.4);
  \draw [->,color=red,thick] (1.8,-2) -- (1.8,-1.6);
  \draw [dashed,->,color=blue,thick] (2.6,-2) -- (2.6,-1.4);
  \draw [->,color=red,thick] (3.4,-2) -- (3.4,-1.6);
  \draw [dashed,->,color=blue,thick] (4.2,-2) -- (4.2,-1.4);
  \draw [->,color=red,thick] (5,-2) --(5,-1.6);
  \end{tikzpicture}
  \caption{Illustration of uniformity of the optimal schedule.
  The dashed and solid arrows denote the transmission instants of the 1st and the 2nd sensor, respectively.
  The schedule $\theta_3$ is the most uniform one and thus better than the other two, regardless of the system parameters and initial states.}
  \label{fig:uniformexample}
\end{figure}
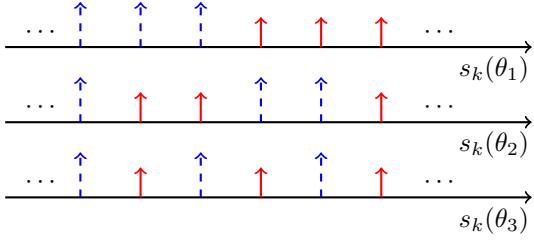

\hlb{Consider $\mathcal{Q}=\{1,2\}$. A periodic schedule that is optimal should not be structured like $\theta_1$ with period $6$ in Fig. \ref{fig:uniformexample}
since one can always construct a better periodic schedule satisfying conditions $(i)$-$(iii)$ in Theorem~\ref{theorem:uniform_transmission} via a procedure of uniformity. For example, $\theta_2$ is more uniform than $\theta_1$ and $\theta_3$ is more uniform than $\theta_2$, resulting in $J(\theta_1)>J(\theta_2)>J(\theta_3)$. Note that the uniformity is a structural property and we can refine the segment in $\theta_1$ as indicated regardless of the system parameters and initial states. This conclusion is in accordance with the result in~\cite{shi2012scheduling}.}
\end{example}

\section{Construction of Suboptimal Schedules}\label{section:suboptimal}
The size of the MDP state space heavily depends on the value of $\Delta_{i}^{\max}~$'s. In the case that there exists $\bb A_i$ with $\rho(\bb A_i))$ far from the spectral radius of other systems, $\Delta_{i}^{\max}$ is extremely large and the state space of the MDP has a large number of states. Even for solving the simple \emph{unichain} MDP, the computational effort per iteration for the policy iteration algorithm is $\sum_{\nu\in\mathcal V}\card{\mathcal A_{\nu}}\card{\mathcal V}+\frac{1}{3}\card{\mathcal V}^3$ \cite{puterman2005markov}. Therefore, how to construct a simple suboptimal schedule is a practical problem.
In this section we introduce several algorithms to search suboptimal periodic schedules, which are computationally simple and independent of the system matrices $\bb A_i$'s. For any suboptimal schedule, we are interested in the performance gap between the optimal and the suboptimal costs. Thus we derive a lower bound of the optimal cost and then manage to obtain an upper bound for the optimality gap.

\subsection{Heuristic Algorithms}
\subsubsection{Maximum Error First (MEF) Algorithm}
We present a simple algorithm, called the Maximum Error First algorithm, which provides a suboptimal schedule to Problem \ref{problem:main-problem-1}.
The idea is to schedule the sensor whose absence causes the maximum trace of estimation error covariance.
\begin{algorithm}
\caption{MEF algorithm}
\label{algorithm:1}
\begin{algorithmic}[1]
\State {\bf Initialization}
\State {\bf Repeat}
\begin{enumerate}
\item Choose $\theta(k)$ such that
\begin{align}
\theta(k)=\argmax_{i\in\mathcal Q} \tr{h_i({\bb P}_{k-1}^{i}(\theta))-{\bb P}_{k-1}^{i}(\theta)}.\label{eqn:6}
\end{align}
\item Update ${\bb P}_{k}^{i}(\theta),~\forall i\in\mathcal Q$.
\end{enumerate}
\end{algorithmic}
\end{algorithm}

\begin{proposition}\label{proposition:MEF}
The schedule $\theta$ searched by the MEF algorithm converges
to be periodic.
\end{proposition}
\begin{pf*}{Proof.}
Since $\rho(\mathbf {A}_i)>1$, the off-duty duration of each sensor is bounded. We can formulate a MDP with finite states. Since the action is uniquely determined by \eqref{eqn:6}, the process is transformed as a Markov chain with transition matrix filled with only $1$ and $0$. Similar to the proof of Theorem \ref{theorem:periodic}, we can conclude that the schedule $\theta$ searched by the MEF algorithm converges to be periodic.
\end{pf*}

\subsubsection{Receding Horizon (RH) Algorithm}
We propose another algorithm, called the Receding Horizon (RH) algorithm in Algorithm \ref{algorithm:2}. The RH algorithm, resembling receding horizon control, computes the optimal scheduling decisions over a finite window of size $\zeta\in \hlb{\mathbb Z_{++}}$ and keeps only the next-step scheduling decision. At the next time instant, it computes a new stretch of scheduling decisions by sliding the window one step forward. Note that the MEF algorithm is a special case of the RH algorithm with $\zeta=1$.
\begin{algorithm}
\caption{RH algorithm}
\label{algorithm:2}
\begin{algorithmic}[1]
\State {\bf Initialization.}
\State {\bf Repeat}
\begin{enumerate}
\item Compute $\theta(k),\ldots,\theta(k-1+\zeta)$ such that they minimize
\begin{align*}
 \sum_{i=k-1}^{k-1+\zeta}\sum_{j=1}^{\card{\mathcal Q}} \left(\tr{{\bb P}_k^j(\theta)}\right).
\end{align*}
\item Keep only $\theta(k)$.
\item Update ${\bb P}_{k}^{i}(\theta),~\forall i\in\mathcal Q$.
\end{enumerate}
\end{algorithmic}
\end{algorithm}

Similar to the MEF algorithm, we have the following result whose proof is omitted.
\begin{proposition}\label{proposition:RH}
The schedule $\theta$ searched by the RH algorithm converges
to be periodic.
\end{proposition}

\begin{remark}By taking the advantage of the asymptotic periodicity property, the MEF and RH algorithms terminate when the schedule reaches a convergence.
It is also convinient for the sensors to store a look-up table of the scheduling
cycle obtained by the MEF or RH algorithm.
\end{remark}

\begin{remark}
\hlb{One can use the property of uniformity to refine the suboptimal schedules like what we do in Example 1. The suboptimal schedules by running numerical algorithms such as the proposed MEF and RH algorithms are often asymptotically periodic. Thus we can apply the procedure of uniformity for each period by interchanging the transmissions across the sensors to improve the performance.}
\end{remark}

\begin{remark}
\hlb{If we set the window size $\zeta$ to be larger than the period of the optimal schedule, we can implement the optimal strategy via the RH algorithm. This property is suitable for a small network for which the optimal period is likely to be small. Since the RH algorithm needs to enumerate all possible $\zeta^n$ strategies, the RH algorithm is unlikely to recover the optimal strategy when $n$ is large subject to the computational capability constraint.}
\end{remark}
\subsection{Theoretical limit of the optimal sensor schedule}
For any suboptimal schedule, we are interested in the optimality gap between the optimal cost induced by an optimal schedule and the suboptimal cost induced by the proposed suboptimal schedule. In the rest of this section, we present a lower bound for the optimal cost
$J(\theta^*)$ by constructing some artificial schedules.
Noting that the transmissions of each individual sensor should be allocated as uniformly as possible from Theorem \ref{theorem:uniform_transmission}, we
consider a class of artificial schedules $\pi:=\{\pi(k)\}$, $\pi(k):\mathbb { Z}\mapsto 2^{\mathcal Q}$, which allow multiple sensors to transmit at the same time and which
make scheduling decisions according to time $k$ as well.
\hlb{Let $\Pi$ be the set of all such feasible artificial schedules.}
For an artificial schedule $\pi$, we denote
as $s_k(\pi)\in 2^{\mathcal{Q}}$ the scheduling decision at time
$k$ and denote as $\mathcal{S}_i(\pi,T)$ the set of all transmissions of the $i$th sensor over $[0,T]$. Some mathematical definitions are
generalized from those of $\theta\in\Theta$ if necessary.
Consider a set of periodic artificial schedules, $$\Pi_{\mathscr P}:=\{\pi\in\Pi:\exists L\in\mathbb{Z}_+,
s_{k}(\pi) = s_{k+L}(\pi),\forall k\in \hlb{\mathbb Z_+}\}.$$
Immediately, we have $\Theta\subseteq\Pi$ and
$\Theta_{\mathscr P}\subseteq\Pi_{\mathscr P}$.
For any $\pi \in\Pi_{\mathscr P}$
Denote
\[F_i(\pi):=\lim_{T\to \infty}\frac{1}{T}\card{\mathcal S_i(\pi,T)}\]
as the duty cycle of the $i$th sensor under  $\pi\in\Pi_{\mathscr P}$.
A first step towards finding the theoretical limit of the optimal sensor schedule within $\Theta_{\mathscr P}$ is to consider the optimal schedule within $\Pi_{\mathscr P}$ given a set of duty cycles satisfying
the constraint
$\sum_{i\in\mathcal Q} F_i(\pi)=1$.
We can find such an optimal schedule $\pi_{\dagger}\in\Pi_{\mathscr P}$ by solving the following problem,
\begin{problem} \label{problem:main-problem-2}
\begin{subequations}
\begin{align*}
\min_{\pi \in \Pi_{\mathscr P}}& ~~~~J(\pi):= \lim_{T\rightarrow\infty} \frac{1}{T}\sum_{i\in\mathcal Q} \sum_{k=0}^{T-1} \left(\tr{{\bb P}_k^i(\pi)} \right)\\
\mathrm{s.t.}&~~~ F_i(\pi)=f_i,~ \forall i\in \mathcal{Q},\\
&~~~\sum_{i\in\mathcal Q} f_i=1.
\end{align*}
\end{subequations}
\end{problem}
For an optimal schedule $\pi_\dagger$
to Problem~\ref{problem:main-problem-2}, an immediate observation based on Theorem \ref{theorem:periodic} is that $\pi_\dagger$ uniformly distributes the transmission of the $i$th sensor subjects to some certain duty cycles (since there is no access constraint any more), and the cost can be computed via the following formula:
\begin{align}
&J(\pi_\dagger)=\nonumber\\
&\sum_{i\in\mathcal Q}\left\{ F_i(\pi_\dagger)\hspace{-1mm}\sum_{j=1}^{n}\tr{h^{j-1}({\bb P}_i)}\hspace{-1mm}+\hspace{-1mm}(1\hspace{-1mm}-\hspace{-1mm}
nF_i(\pi_\dagger))\tr{h^{n}({\bb P}_i)}\right\}\label{eqn:15}
\end{align}
where $n:=\max\{j\in\hlb{\mathbb Z_+}:jF_i(\pi_\dagger)\leq 1\}.$

Next we shall investigate how the duty cycles of the sensors
affect the cost of the schedule
by studying how $J(\pi_\dagger)$ relates to $F_i(\pi_\dagger)$'s in~\eqref{eqn:15}.
For simplicity of notations, define a set of
piecewise linear functions $\phi_i:(0,1]\mapsto(0,\infty), i\in\mathcal Q$,
\begin{equation}
\phi_i(z) =c_i(z) z +\tr{h_i^{\beta}({\bb P}_i)},\label{eqn:piecewise-linear-fun}
\end{equation}
where
\begin{align*}
c_i(z) &= -\left\lfloor\frac{1}{z}\right\rfloor \tr{h_i^{\left\lfloor\frac{1}{z}\right\rfloor}({\bb P}_i)}+\sum_{j=0}^{\left\lfloor\frac{1}{z}\right\rfloor-1}
\tr{h_i^j({\bb P}_i)},\\
\beta &=\left\lfloor\frac{1}{z}\right\rfloor\in \hlb{\mathbb Z_{++}}.
\end{align*}
Then we have $J(\pi_\dagger)=\sum_{i\in\mathcal Q}\phi_i(F_i(\pi_\dagger))$. The  piecewise linearity and convexity of $\phi_i(z)$ are given in the following lemma.
\begin{lemma}\label{lemma:convex}
For any $i\in\mathcal Q$, the function $\phi_i(z)$ defined in~\eqref{eqn:piecewise-linear-fun} is continuous, convex and piecewise linear, \ie, any segment between points $1/\alpha$ and $1/(\alpha+1)$, where $\alpha\in \hlb{\mathbb Z_+}$ and $\alpha\geq \card{\mathcal Q}$, is affine.
\end{lemma}
\begin{pf*}{Proof.}
The piecewise linearity and continuity are easy to see. We
shall verify the convexity by showing that
$c_i(z)$ is a nondecreasing function of $z$. For $z_2\leq z_1$, letting $\beta_2=\left\lfloor\frac{1}{z_2}\right\rfloor$
and  $\beta_1=\left\lfloor\frac{1}{z_1}\right\rfloor$, we have
\begin{align*}
&c_i(z_2)-c_i(z_1)\\
=& -\beta_2\tr{h_i^{\beta_2}({\bb P}_i)}+\beta_1\tr{h_i^{\beta_1}({\bb P}_i)}+\sum_{j=\beta_1}^{\beta_2-1}\tr{h^j({\bb P}_i)}\\
=& -\beta_1\tr{h_i^{\beta_2}({\bb P}_i)-h_i^{\beta_1}({\bb P}_i)}\\
&~~~~~~~~~~~~~~~~~~~~~~~~~+\sum_{j=\beta_1}^{\beta_2-1}
\tr{h_i^j({\bb P}_i)-h_i^{\beta_2}({\bb P}_i)}\leq 0.
\end{align*}
The last inequality is from Lemma \ref{lemma:supporting-lemma}, which
completes the proof.
\end{pf*}

\hlb{The above lemma can help us find a lower bound for
$J(\theta^*)$ by solving  a convex programming optimization
problem equivalent to Problem \ref{problem:main-problem-2}, which is formulated as follows:
\begin{problem}\label{problem:cvx}
\begin{align*}
\min_{\{f_i\}} ~~&\sum_{i\in\mathcal Q}\phi_i(f_i),\\
\rm{s.t.}~~& \sum_{i\in\mathcal Q}f_i=1,\\
& \frac{1}{\Delta_i^{\max}}\leq f_i\leq 1 - \sum_{j\neq i,j\in\mathcal Q}\frac{1}{\Delta_j^{\max}},~\forall i\in\mathcal{Q}.
\end{align*}
\end{problem}
Then we are in the position to give a lower bound of the
optimal cost for Problem \ref{problem:main-problem-1}.
\begin{proposition}\label{proposition:lower_bound}
For an optimal schedule $\theta^*\in\Theta_{\mathscr P}$ of
Problem~\ref{problem:main-problem-1}, a lower bound of $J(\theta^*)$ is given by
\[J(\theta^*)\geq J(\pi^*),\]
where $\pi^*\in\Pi_{\mathscr P}$ is an optimal solution to
Problem~\ref{problem:main-problem-2} with the duty cycles
$F_i(\pi^*)$'s obtained by solving Problem~\ref{problem:cvx}.
\end{proposition}
The fact is due to $\Theta_{\mathscr P}\subseteq
\Pi_{\mathscr P}$.
Note that the inequality constraint in Problem \ref{problem:cvx} gives a tighter lower bound than $0<f_i<1$ since the feasible duty cycle of the $\theta^*$ is subject to the constraint in Problem \ref{problem:cvx}.}
In the following we also give a structural description of
the solution to Problem \ref{problem:cvx}.
\begin{lemma}\label{lemma:joint_point}
{There exists a solution $(f_1,\ldots,f_{|\mathcal Q|})$
to Problem \ref{problem:cvx} such that all
$j\in\mathcal Q$ but at most one point satisfy
 $1/f_j \in \hlb{\mathbb Z_{++}}$.}
\end{lemma}
\begin{pf*}{Proof.}
For an solution $(f_1,\ldots,f_{|\mathcal Q|})$ to Problem~\ref{problem:cvx}, without loss of
generality, we assume that
there exist $i,j\in\mathcal Q$ and
$z_1,z_2\in\hlb{\mathbb Z_{++}}$ such that
\[f_i\in(\frac{1}{z_1},\frac{1}{z_1+1}),
\hbox{~and~}f_j\in(\frac{1}{z_2},\frac{1}{z
_2+1}),\]
and that  $c_i(f_i)\leq c_j(f_j)$. Then,
\begin{equation}\label{eqn:16}
\sum_{\ell\in\mathcal Q}\phi_\ell(f_\ell) \geq \phi_i(f_i+\delta)+\phi_j(f_j-\delta)+ \sum_{\ell\neq i,j}\phi_\ell(f_\ell),
\end{equation}
where $\delta=\max\left\{1/(f_1+1)-f_i,f_j-1/f_2\right\}$.
Then~\eqref{eqn:16} shows that there must exist a solution $(f_1,\ldots,f_i+\delta,\ldots,
f_j-\delta,\ldots,f_{|\mathcal Q|})$
at least as good as $(f_1,\ldots,f_{|\mathcal Q|})$.
Finally note that either $f_i+\delta$ or $f_j-\delta$ takes the form $1/\alpha$ for some
$\alpha\in\hlb{\mathbb Z_{++}}$, which completes the proof.
\end{pf*}

The optimal duty cycles can be even used to search an optimal solution to Problem \ref{problem:main-problem-1} as long as we can construct a periodic schedule $\theta\in\Theta_{\mathscr P}$ with the optimal duty cycles obtained by solving Problem \ref{problem:cvx} and
$\theta$ is of uniformity. The result is formally summarized as follows.
\begin{proposition}\label{proposition:construction}
If a schedule $\theta\in\Theta_{\mathscr P}$ satisfies
\begin{enumerate}
  \item[(i).] Optimal duty cycles: $F_i(\theta)$'s are a solution to Problem~\ref{problem:cvx}, and
  \item[(ii).] Uniformity: let $L$ be the period of $\theta$,
  \[
    \left|\bb{d}^i(\theta,L)[j_1]
    -\bb{d}^i(\theta,L)[j_2]\right|\leq 1,\]
    for any $j_1,j_2\in\{1,\ldots,\sigma_{i}(L)\}$ and
    any $i\in\mathcal Q$;
\end{enumerate}
then $\theta$ is an optimal solution to Problem \ref{problem:main-problem-1}.
\end{proposition}
To see how Proposition~\ref{proposition:construction}
is used for searching an optimal solution
to Problem \ref{problem:main-problem-1}, let us take a two-sensor case as an example.
First we must have $f_1=1/z_0$ and $f_2=(z_0-1)/z_0$ for some
$z_0\in\mathbb{Z}_+$ as a solution to
Problem~\ref{problem:cvx}. It can be seen a schedule satisfying $(i)$ and $(ii)$ in Proposition \ref{proposition:construction} can be constructed, where at each period a single transmission of the $1$st sensor
followed by $z_0-1$ consecutive transmissions of the $2$nd sensor.
Such a construction for a multiple-sensor case, however, is not trivial. For instance, for a three-sensor case, where
$f_1=1/6,~f_2=1/3,~f_3=1/2$ form a solution to Problem~\ref{problem:cvx}, it is impossible to construct a schedule satisfying $(ii)$ of Proposition \ref{proposition:construction}. In other words, the one-time-one-transmission constraint may severely obstruct uniform scheduling given optimal duty cycles obtained by solving Problem \ref{problem:cvx}. The feasibility of the construction and implementable algorithms are two interesting open problems.

%
%

\section{Discussion on Scheduling Stable Systems}
Though we have assumed that ${\bb A}_i$'s are unstable since estimation performance may become unpredictable if a process is left unattended for a long time, this question still makes sense: what about scheduling multiple systems, some of which are stable systems?

If there are stables systems involved, Theorem \ref{theorem:maximum_gap} on the boundedness of off-duty durations may not hold any more. Let us look at an example. It is well known that the estimation error covariance for a stable system $(A_i,Q_i)$ is the solution $\bb P_i^{Lyn}$ to the Lyapunov equation $X=A_iXA_i^{\top}+Q_i$. \hlb{If for any $j$th sensor, where $j\neq i$, the following inequality holds}:
\[
\tr{h_j^\ell(h_j(\bb P_j) - \bb P_j)} \geq \tr{\bb P_i^{Lyn} - \bb P_i}, \forall l\in \hlb{\mathbb Z_+},
\]
then the $i$th sensor will never be scheduled since its transmission helps decrease the average error covariance less than that of any other sensor does in all circumstances. Therefore, a MDP formulation may not be appropriate for this case due to the unclear state space.

However, the property of uniformity in Theorem \ref{theorem:uniform_transmission} still holds for the optimal schedule regardless of the choices of $A$'s. Thus we are still able to use the heuristics algorithms in Section \ref{section:suboptimal} to find a suboptimal solution. The optimality gap can be quantified by using Proposition \ref{proposition:lower_bound}.

\section{Numerical Examples}
\subsection{An Illustrating Example for a Small Sensor Network}
\begin{figure}
  \centering
  \includegraphics[width=3.4in]{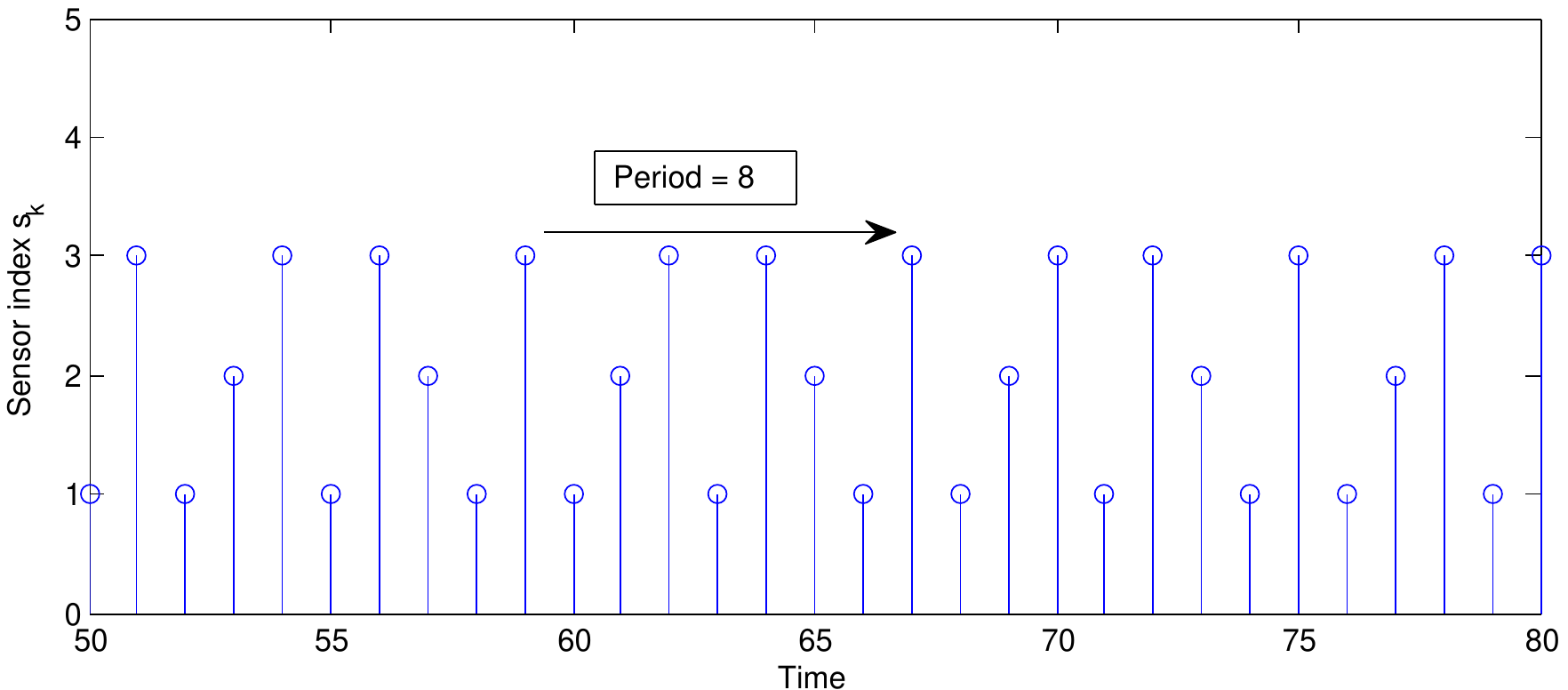}\\
  \caption{Realization of the action trajectory for an optimal schedule converges to be periodic. {The matchsticks with different lengths represent network accesses from different sensors.}
  The period is $8$.}\label{fig:realization}
\end{figure}

In this section we give some numerical simulations to illustrate the main theoretical results. Consider three sensors monitoring three different linear systems. The system models are given as follows:
\begin{align*}
\bb A_1&=\begin{bmatrix}1.1&1.2\\0&1\end{bmatrix},\bb A_2=\begin{bmatrix}1.5&1.2 \\0&1\end{bmatrix},\bb A_3=\begin{bmatrix}1.9&1.2 \\0&1\end{bmatrix},\\
\bb C_1 &= \begin{bmatrix}1&1 \end{bmatrix},\bb C_2 = \begin{bmatrix}1&1\end{bmatrix},\bb C_3 = \begin{bmatrix}1&1\end{bmatrix},\\
\bb Q_1&=\begin{bmatrix}26&0 \\0&5\end{bmatrix},\bb Q_2=\begin{bmatrix}3&0 \\0&1\end{bmatrix},\bb Q_3=\begin{bmatrix}4& 0\\0&32\end{bmatrix},\\
\bb R_1 &= 1,\bb R_2 = 1,\bb R_3 = 1.
\end{align*}
First we numerically obtain $\Delta_1^{{\rm max}}=32,~\Delta_2^{{\rm max}}=17$ and $\Delta_3^{{\rm max}}=7$, the bound of the off-duty duration for each sensor in an optimal schedule. Thereafter we can formulate an MDP problem with the number of states $747$. We use the policy iteration \cite[Chapter 9.2]{puterman2005markov} to search the optimal schedule. We initiate the iteration with a random allowable decision rule. After $7$ iterations, the algorithm reaches the convergence and the computational time is $122.10s$. The optimal cost function is $144.0$. A realization of the action trajectory at steady state of the optimal schedule is depicted in Fig. \ref{fig:realization}.
An optimal schedule is shown to be
\begin{align*}
&s_{8k-7}(\theta)=3,s_{8k-6}(\theta)=1,s_{8k-5}(\theta)=2,s_{8k-4}(\theta)=3,\\
&s_{8k-3}(\theta)=1,s_{8k-2}(\theta)=3,s_{8k-1}(\theta)=2,s_{8k}(\theta)=1,
\end{align*}
where we can notice the uniformity of the transmissions of each sensor.

\hlb{For comparison, we also find the suboptimal schedules using the MEF algorithm and the RH algorithm with different window sizes \footnote{
All the algorithms in this section are tested in MATLAB R2013b on a Windows PC platform with i7-4600U CPU (2.10GHz, 4 CPUs) and 8G RAM.}. The RH-$\zeta$ stands for the RH algorithm with the window size of $\zeta$. The computation time and the value of cost function for all algorithms are listed in Table \ref{table:comparison1}. As a reference, we also show the lower bound by solving Problem \ref{problem:cvx}, abbreviated as LB. Notice that searching the optimal schedule by solving an MDP problem is the most time consuming. The MEF runs very fast but the performance is rather poor compared with other algorithms. Note that the cost value of RH-$5$ collides with the optimal value but generally it cannot guarantee the optimality unless the window size is larger than the period of the optimal schedule.}

\begin{table} 
\centering
\begin{tabular}{cccccc}
  \toprule
  & MDP & MEF & RH-$2$ & RH-$5$ & LB\\
  & \hspace{-3pt}(optimal)\hspace{-3pt} &  &  &  & \\
  CPU  time(s)& 122.1 & 0.5 & 0.7 & 5.6 & -\\
  Cost value& 144.0 & 161.3 & 145.4 & 144.0 & 140.1\\
  \bottomrule
\end{tabular}
\vspace{5pt}
\caption{Comparison among the optimal schedule and suboptimal schedules.}
\label{table:comparison1}
\end{table}

\subsection{Computational Disaster for solving an MDP Problem}
\begin{table}[]
\centering
\begin{tabular}{cccccc}
  \toprule
    & MDP & MEF & RH-$2$ & RH-$5$ & LB\\
  & \hspace{-3pt}(optimal)\hspace{-3pt} &  &  &  & \\
  CPU time(s)& 660.6 & 0.7 & 0.8 & 6.8 & -\\
  Cost value & 116.1 & 121.4 & 116.1 & 116.1 & 109.5\\
  \bottomrule
\end{tabular}
\vspace{5pt}
\caption{Comparison among the optimal schedule and suboptimal schedules.}
\label{table:comparison2}
\end{table}

In this section we show that the computational time of solving an MDP heavily relies on the size of state space. If we consider the following system matrices with the same $\bb C_i,\bb Q_i,\bb R_i$ as above:
\begin{align*}
\bb A_1&=\begin{bmatrix}1.001&1.2\\0&1\end{bmatrix},\bb A_2=\begin{bmatrix}1&0.8 \\0&1.001\end{bmatrix},\bb A_3=\begin{bmatrix}1.2&1 \\0&1.1\end{bmatrix},
\end{align*}
we obtain $\Delta_1^{{\rm max}}=22,~\Delta_2^{{\rm max}}=45$ and $\Delta_3^{{\rm max}}=7$, and the state space becomes $1278$. In Table \ref{table:comparison2}, we compare the computational time and cost value across different algorithms. Notice that the computational time for solving an MDP is as long as $660.60$s, roughly $5$ times of the previous example, but the state space increases only by $71.1\%$. On the contrary, the time varies little for suboptimal schedules compared with the previous example in Table \ref{table:comparison1}. 

\hlb{Consider a large-scale network of $15$ sensors instead. The parameters are given as 
\begin{align*}
\bb A_1&=\begin{bmatrix}1.07& 0.07\\0&1.07\end{bmatrix},\bb A_2=\begin{bmatrix}1.13&0.13 \\0&1.13\end{bmatrix},\bb A_3=\begin{bmatrix}1.2&0.2 \\0&1.2\end{bmatrix},\\
\bb A_4&=\begin{bmatrix}1.27& 0.27\\0&1.27\end{bmatrix},\bb A_5=\begin{bmatrix}1.33&0.33 \\0&1.33\end{bmatrix},\bb A_6=\begin{bmatrix}1.4&0.4 \\0&1.4\end{bmatrix},\\
\bb A_7&=\begin{bmatrix}1.47& 0.47\\0&1.47\end{bmatrix},\bb A_8=\begin{bmatrix}1.53&0.53 \\0&1.53\end{bmatrix},\bb A_9=\begin{bmatrix}1.6&0.6 \\0&1.6\end{bmatrix},\\
\bb A_{10}&=\begin{bmatrix}1.67& 0.67\\0&1.67\end{bmatrix},\bb A_{11}=\begin{bmatrix}1.73&0.73 \\0&1.73\end{bmatrix},\bb A_{12}=\begin{bmatrix}1.8&0.8 \\0&1.8\end{bmatrix},\\
\bb A_{13}&=\begin{bmatrix}1.87& 0.87\\0&1.87\end{bmatrix},\bb A_{14}=\begin{bmatrix}1.93&0.93\\0&1.93\end{bmatrix},\bb A_{15}=\begin{bmatrix}2&1 \\0&2\end{bmatrix},
\end{align*}
\begin{align*}
 \bb C_1&=\begin{bmatrix}-0.93& 0.07\\0.07&-1.93\end{bmatrix},\bb C_2=\begin{bmatrix}-0.87&0.13 \\0.13&-1.87\end{bmatrix},\\\bb C_3&=\begin{bmatrix}-0.8&0.2 \\0.2&-1.8\end{bmatrix},
\bb C_4=\begin{bmatrix}-0.73& 0.27\\0.27&-1.73\end{bmatrix},\\\bb C_5&=\begin{bmatrix}-0.67&0.33 \\0.33&-1.67\end{bmatrix},\bb C_6=\begin{bmatrix}-0.6&0.4 \\0.4&-1.6\end{bmatrix},\\
\bb C_7&=\begin{bmatrix}-0.53& 0.47\\0.47&-1.53\end{bmatrix},\bb C_8=\begin{bmatrix}-0.47&0.53 \\0.53&-1.47\end{bmatrix},\\\bb C_9&=\begin{bmatrix}-0.4&0.6 \\0.6&-1.4\end{bmatrix},
\bb C_{10}=\begin{bmatrix}-0.33& 0.67\\0.67&-1.33\end{bmatrix},\\
\bb C_{11}&=\begin{bmatrix}-0.27&0.73 \\0.73&-1.27\end{bmatrix},\bb C_{12}=\begin{bmatrix}-0.2&0.8 \\0.8&-1.2\end{bmatrix},\\
\bb C_{13}&=\begin{bmatrix}-0.13& 0.87\\0.87&-1.13\end{bmatrix},\bb C_{14}=\begin{bmatrix}-0.07&0.93 \\0.93&-1.07\end{bmatrix},\\ \bb C_{15}&=\begin{bmatrix}0&1 \\1&-1\end{bmatrix},
\end{align*}
\begin{align*}
\bb Q_i&=10^{-6}\I,\forall i\in[1,15],\\
\bb R_1 &= 1.07\times 10^{-6}\I,\bb R_2 = 1.13\times 10^{-6}\I,\\
\bb R_3 &= 1.2\times 10^{-6}\I,
\bb R_4 = 1.27\times 10^{-6}\I,\\
\bb R_5 &= 1.33\times 10^{-6}\I,\bb R_6 = 1.4\times 10^{-6}\I,\\
\bb R_7 &= 1.47\times 10^{-6}\I,\bb R_8 = 1.53\times 10^{-6}\I,\\
\bb R_9 &= 1.6\times 10^{-6}\I,
\bb R_{10} = 1.67\times 10^{-6}\I,\\
\bb R_{11} &= 1.73\times 10^{-6}\I,\bb R_{12} = 1.8\times 10^{-6}\I,\\
\bb R_{13} &= 1.87\times 10^{-6}\I,\bb R_{14} = 1.93\times 10^{-6}\I,\\
\bb R_{15} &= 2\times 10^{-6}\I.
\end{align*}
The bounds of the off-duty duration for all sensors are $163$, $163$, $163$,   $163$, $163$, $163$, $147$, $131$, $119$, $108$, $102$, $96$, $91$,  $86$, $43$. The size of the state space is roughly $10^{31}$. Solving an MDP problem is formidable in this case. We design four suboptimal schedules by using MEF, RH-$2$, RH-$3$ and RH-$5$, respectively. In Table \ref{table:comparison3}, we list all the cost value and the computation time, and the lower bound for the optimal value as a reference.
\begin{table}[]
\centering
\begin{tabular}{cccccc}
  \toprule
    & MEF & RH-$2$ & RH-$3$ & RH-$5$ & LB\\
  CPU time& 7.1s & 87.9s & 0.5hr  & 6.4hr & -\\
  Cost value & 47.2 & 45.0 & 43.5 & 40.3 &22.1\\
  \bottomrule
\end{tabular}
\vspace{5pt}
\caption{Comparison among the suboptimal schedules for a large network.}
\label{table:comparison3}
\end{table}
}

To summarize, when the large state space of the MDP formulation prevents the efficient computation of the optimal solution, the RH algorithm is a good candidate to search a suboptimal solution and the optimality gap can be bounded by the difference between the suboptimal cost value and the lower bound of the optimal cost value by solving Problem \ref{problem:cvx}.

\subsection{Optimal Schedule Construction}
In some cases, we can directly construct the optimal schedule from the solution to Problem \ref{problem:cvx}. For example, consider the following set of systems
{\small
\begin{align*}
\bb A_1&=\begin{bmatrix}1.18&1.31\\0&2.66\end{bmatrix},\bb A_2=\begin{bmatrix}1.95&1.99 \\0&2.40\end{bmatrix},\bb A_3=\begin{bmatrix}2.65&1.75 \\0&1.45\end{bmatrix},\\
\bb C_1 &= \begin{bmatrix}1&1 \end{bmatrix},\bb C_2 = \begin{bmatrix}1&1\end{bmatrix},\bb C_3 = \begin{bmatrix}1&1\end{bmatrix},\\
\bb Q_1&=\begin{bmatrix}0.1&0 \\0&9.51\end{bmatrix},\bb Q_2=\begin{bmatrix}0.81&0 \\0&4.78\end{bmatrix},\bb Q_3=\begin{bmatrix}0.38& 0\\0&5.11\end{bmatrix},\\
\bb R_1 &= 1,\bb R_2 = 1,\bb R_3 = 1.
\end{align*}
}
we find the optimal duty cycle $f_1=0.5,f_2=0.25$ and $f_3=0.25$ for each sensor. Then we can construct a collision-free policy directly, \ie, for all $k\in \hlb{\mathbb Z_+}$
$$s_{4k-3}(\theta)=1,s_{4k-2}(\theta)=2,s_{4k-1}(\theta)=1,s_{4k}(\theta)=3,$$
and it proves to be optimal by Proposition \ref{proposition:construction}.

\section{Concluding Remarks}\label{section:conclusion}
We considered the optimal scheduling problem for multiple sensors monitoring different linear dynamical systems but sharing one common link to the remote estimator. At each time only one sensor can communicate with the remote estimator. To manage the network access, we restricted our attention to
 {time-based} sensor transmission schedules.
We first presented a necessary condition for optimality provided that the spectral radii of any two system matrices are not equal, greatly curtailing the feasible solution space
without loss of performance,
and then formulated a finite-state MDP problem to search the optimal schedule. Next we showed the asymptotic periodicity and uniformity properties of an optimal schedule.
The computational complexity of solving an MDP problem, however, is formidable in some cases. Thus we proposed two simple suboptimal schedules to bypass the computational burden and also quantified the optimality gap between the optimal cost and the suboptimal ones.

The two-sensor scheduling result in \cite{shi2012scheduling} is a special case of this work, while an optimal solution to scheduling over two sensors is complicated and can no longer be explicitly written. Better necessary conditions like Theorem \ref{theorem:maximum_gap} are essential for reducing the feasible solution space and thus enhancing the performance of the numerical algorithm for searching the optimal solution. There are some interesting open problems. For example, a tighter bound for the off-duty duration of each sensor is desired to decrease the number of the MDP states. Inspired by Proposition \ref{proposition:construction}, how to directly construct an optimal or suboptimal schedule is promising but still challenging since in some cases the construction is a combinatorial optimization problem.

\bibliographystyle{plainnat}
\bibliography{automatica_references}

\end{document}